\documentclass[12pt,preprint]{aastex}

\usepackage{natbib}
\usepackage{pdflscape}

\def\gcc{\hbox{\rm\hskip.35em  g cm}$^{-3}$}
\def\rads{\hbox{\rm\hskip.35em  rad s}$^{-1}$}
\def\cms{\hbox{\rm\hskip.35em  cm s}$^{-1}$}

\def\eg{{\it e.g.}}
\def\lap{\hbox{${_{\displaystyle<}\atop^{\displaystyle\sim}}$}}
\def\gap{\hbox{${_{\displaystyle>}\atop^{\displaystyle\sim}}$}}

\newcommand{\ubf}       {\mbox{\boldmath$u$}}
\newcommand{\vbf}       {\mbox{\boldmath$v$}}
\newcommand{\vsbf}       {\mbox{\boldmath$v_s$}}
\newcommand{\kappabf}       {\mbox{\boldmath$\kappa$}}
\newcommand{\omegabf}       {\mbox{\boldmath$\omega$}}
\newcommand{\fbf}       {\mbox{\boldmath$f$}}
\newcommand{\Fbf}       {\mbox{\boldmath$F$}}

\def\eg{{\it e.g.}}
\def\lap{\hbox{${_{\displaystyle<}\atop^{\displaystyle\sim}}$}}
\def\gap{\hbox{${_{\displaystyle>}\atop^{\displaystyle\sim}}$}}

\usepackage{graphicx}
\usepackage{subfigure}
\begin{document}

\title{THERMALLY-ACTIVATED POST-GLITCH RESPONSE OF
THE NEUTRON STAR INNER CRUST AND CORE. I: THEORY}

\author{Bennett Link}
\affil{Department of Physics, Montana State University, Bozeman,  MT
59717, USA:}
\email{link@physics.montana.edu}

\date{\today}
\begin{abstract}
Pinning of superfluid vortices is predicted to prevail throughout much
of a neutron star. Based on the idea of Alpar et al., I develop a
description of the coupling between the solid and liquid components of
a neutron star through {\em thermally-activated vortex slippage}, and
calculate the the response to a spin glitch.  The treatment begins
with a derivation of the vortex velocity from the vorticity equations
of motion. The activation energy for vortex slippage is obtained from
a detailed study of the mechanics and energetics of vortex motion. I
show that the ``linear creep'' regime introduced by Alpar et al. and
invoked in fits to post-glitch response is not realized for physically
reasonable parameters, a conclusion that strongly constrains the
physics of post-glitch response through thermal activation. Moreover,
a regime of ``superweak pinning'', crucial to the theory of Alpar et
al. and its extensions, is probably precluded by thermal
fluctuations. The theory given here has a robust conclusion that can
be tested by observations: {\em for a glitch in spin rate of magnitude
$\Delta\nu$, pinning introduces a delay in the post-glitch response
time}. The delay time is $t_d=7 (t_{sd}/10^4\mbox{
yr})((\Delta\nu/\nu)/10^{-6})$ d where $t_{sd}$ is the spin-down age;
$t_d$ is typically weeks for the Vela pulsar and months in older
pulsars, and is independent of the details of vortex
pinning. Post-glitch response through thermal activation cannot occur
more quickly than this timescale. Quicker components of post-glitch
response as have been observed in some pulsars, notably, the Vela
pulsar, cannot be due to thermally-activated vortex motion but must
represent a different process, such as drag on vortices in regions
where there is no pinning. I also derive the mutual friction force for
a pinned superfluid at finite temperature for use in other studies of
neutron star hydrodynamics.
\end{abstract}


\section{Introduction}

The problem of the origin of spin glitches in neutron stars has
remained unsolved since the first glitch was observed in the Vela
pulsar in 1969. This difficult problem is of considerable interest, as
an understanding of the glitch phenomenon would offer insights into
both the dynamical and ground-state properties of matter above nuclear
density. To make progress on the glitch problem, it is crucial to
identify which components of the liquid interior corotate with the
crust, which store and release the angular momentum that drives glitches,
and which produce the observed post-glitch response that occurs over days to
years (for examples, see
\citealt{lohsen75,lyne87,cordes_etal88,sl96,wang_etal00,wang_etal01,espinoza_etal11}).
A more complete understanding of the processes that regulate
post-glitch response would be very useful for elucidating the
dynamical properties of the neutron star interior. Fortunately, it is
possible to address this problem without knowledge of the basic
instability that causes glitches, and progress is possible 
with minimal assumptions.

The neutron star interior was predicted long ago to contain superfluid
neutrons and superconducting protons (\citealt{migdal59,gk65}; for
recent reviews, see \citealt{dh03} and \citealt{ch08}). As in all
rotating neutral superfluids, the neutron superfluid is threaded by an
array of quantized vortices whose arrangement determines the angular
momentum of the superfluid, and whose motion determines the torque
exerted on the charged components of the star and the crust. In
laboratory liquid helium, vortices {\em pin} to bumps on the bottom of
the vessel, trapping the system in metastable rotational states. As
the vessel is spun down, jumps in its spin rate are observed
\citep{tt80}, much like the glitches observed in neutron stars, as
vortices unpin and the superfluid transfers angular momentum to the
vessel.

Vortex pinning is predicted to occur throughout much of a neutron
star, with the vortex lines pinning to sites along their length. In
the inner crust, where a neutron superfluid coexists with the ionic
lattice, vortices interact with nuclei with energies of $\sim 1$ MeV
per nucleus \citep{alpar77,eb88,dp06,abbv07}. This interaction pins
the vortex lattice to the nuclear lattice, as confirmed by the
simulations of \citet{link09}. In the outer core, the neutron flow
around the vortices that thread the rotating superfluid entrains a
proton mass current, strongly magnetizing the vortices as shown in the
seminal work of
\citet{als84}. Here the protons are expected to form a type II
superconductor, with the magnetic flux arranged in flux tubes that are
frozen to the highly-conductive charged component of the fluid. The
proton superconductor rotates as a rigid body, without forming
vortices. The vortices of the neutron superfluid pin to the flux
tubes, primarily through a magnetic interaction, with pinning energies
as high as $\sim 100$ MeV per vortex-flux tube junction
\citep{srinivasan_etal90,jones91,mendell91a,chau_etal92,rzc98,link12b}. While
pinning energies remain rather uncertain, the conclusion that
vortices pin to inner-crust nuclei and outer-core flux tubes 
is on solid ground.

The nature of the glitch instability is a problem under active
research, and many possibilities have been suggested. 
As originally suggested by \citet{ai75}, glitches
could represent sudden unpinning of many vortices; as liberated vortices
move under dissipation, a spin-up torque is exerted on the crust plus
charged components. In this connection, glitches could arise through a
vortex avalanche from a critical state
\citep{cheng_etal88,mpw08}. Other possibilities include 
increased frictional coupling from sudden heating of the crust
\citep{greenstein79a,le96} such as by a starquake, crust failure under
superfluid stresses \citep{ruderman91c}, or interactions of magnetized
flux tubes in the outer core with the London current near the core
boundary \citep{sc99}.  \citet{pizzochero11} has proposed that
glitches result from the motion of a vortex sheet into a region where
pinning is not sustainable. Alternatively, glitches might represent
transitions between different states of superfluid turbulence
\citep{peralta_etal06,pm09,ga09}, though these studies do not account
for vortex pinning which is likely to play an important role in the
dynamics. Recent work has shown that pinning is
hydrodynamically unstable in both the inner crust
\citep{link12a} and the outer core
\citep{link12b}; further work is need to see how the
instability saturates, and if the system turns
turbulent. 

Vortex pinning decreases the rotational coupling between the portions of
the superfluid in which there is pinning and the rest of the star. If
vortices remained perfectly pinned between glitches, the decoupling
would be complete, and the superfluid would exert no torque on the 
charged components of the star (the crust, electrons, and core proton
fluid). At finite temperature, though, vortices move through thermal
activation over their pinning barriers, with an associated coupling
time that is generally far longer than without pinning.  Post-glitch
response has been ascribed to internal torques exerted on the crust as
the pinned superfluid responds to the glitch through
thermally-activated vortex motion, as in the ``vortex creep theory''
of
\citet{alpar_etal84a}, hereafter AAPS, that has 
been substantially developed and used to fit pulsar
timing data
\citep{alpar_etal84b,alpar_etal85,alpar_etal86,alpar_etal88,alpar_etal89,alpar_etal93,alpar_etal94,alpar_etal96}.

In vortex creep theory there are two dynamical regimes: the
``non-linear creep'' regime introduced by AAPS, and the ``linear creep''
regime introduced by
\citet{alpar_etal89}. In the ``non-linear creep'' regime, the system is
always near the threshold for vortex unpinning. Post-glitch response
is determined by two timescales, the {\em decoupling time} $t_d$ and
the {\em recoupling time} $t_r$, and post-glitch response depends
non-linearly on the size of the glitch.\footnote{In AAPS, $t_d$ is
denoted $t_o$, and is called the ``offset time", while $t_r$ is
denoted $\tau$, and is called the ``relaxation time''.} Typical
response in the ``nonlinear" creep regime is depicted in
Figure \ref{response_resid}; the system remains out of equilibrium for a
time $t_d$, before recovering over a time $t_r$, with a total response
time of $t_d+t_r$. By contrast, in the ``linear creep" regime
post-glitch response depends linearly on the size of the glitch, and
consists of simple exponential recovery over one timescale rather than
two. The ``linear creep" regime is generally associated with regions
in which pinning is very weak, termed ``superweak pinning" by AAPS.

The "non-linear creep regime" does admit linear response in the limit
$t_r>>t_d$. The important dynamical difference between the
two regimes is that the total recovery time can be very short in the
``linear creep" regime, but always exceeds $t_d$ in the ``non-linear" regime.

As vortex creep theory became more developed, terms were introduced in
the crust response to account for internal torques from different
parts of the star, vortex depletion regions, and crust cracking; see,
\eg, \citet{alpar_etal96}. Prompt exponential response is introduced
by assuming the existence of regions that are in the ``linear creep''
regime. Fits to data introduce many free parameters, and 
do not offer conclusions with which the theory can be refuted. In
particular, the ``linear creep" regime, which gives simple exponential
response independent of glitch magnitude, is not well-constrained by
observations as it is observationally indistinguishable from vortex
motion in a drag regime in which there is no pinning at all (see 
 Section \ref{dragregime}).  

Inspired by the work of AAPS, the purpose of this paper is to present
a comprehensive theory of thermally-activated vortex motion in the
inner crust and the outer core with which to interpret existing and
future glitch data and, in particular, to make {\em falsifiable
predictions}. The theory developed here, which I call
``thermally-activated vortex slippage'' or ``vortex slippage'', to
distinguish it from the ``vortex creep theory'' of AAPS, begins with a
derivation of the vortex velocity from the vorticity equations of
motion, an ingredient that is lacking from the vortex creep theory of
AAPS and its extensions. The activation energy for vortex slippage is
obtained from a detailed study of the mechanics and energetics of
vortex motion.  A key conclusion of this paper is that the ``linear
creep" regime invoked in vortex creep theory is not realized for
physically reasonable parameters. The ``linear creep" regime relies on
vortex motion anti-parallel to the Magnus force to be nearly equal to
vortex motion parallel to the Magnus force. Here I show that the
anti-parallel motion is so strongly suppressed by the large vortex
self-energy as to be negligible.  Fits to post-glitch data that invoke
this regime \citep{alpar_etal93,alpar_etal94,alpar_etal96} should
therefore be reassessed. Without a ``linear creep'' regime of vortex
response, the problem of vortex motion through thermal activation
becomes strongly constrained; effectively, only the ``non-linear"
regime of thermally-activated vortex motion remains.  A second
important conclusion is that ``superweak pinning", which plays a key
role in vortex creep theory, probably does not occur but is eliminated
by thermal fluctuations.

The other conclusions of this paper are in {\em qualitative} agreement
with the ``non-linear" regime studied by AAPS. In particular:

\begin{enumerate}

\item If pinning occurs, post-glitch response through
thermally-activated vortex slippage consists of two distinct
phases. First, the pinned superfluid is {\em decoupled} from the crust
over a timescale $t_d$, during which the external torque acts on
a lower moment of inertia, and the crust spins down at a greater
rate. After a time $t_d$, a region in which there is pinning
will recouple over a timescale $t_r$. The total recovery time from
the glitch is $t_d+t_r$. Post-glitch response generally depends {\em
non-linearly} on the initial conditions. 

\item For a glitch in the spin rate of magnitude $\Delta\nu$, the
decoupling time is 
\begin{equation}
t_d\simeq 7\,\left(\frac{t_{sd}}{10^4\mbox{ yr}}\right)
\left(\frac{\Delta\nu/\nu}{10^{-6}}\right)
\mbox{ days,}
\end{equation}
where $t_{sd}$ is the spin-down age. This timescale represents a {\em
lower limit} for the duration of post-glitch response. 

\item If the recoupling time $t_r$
satisfies $t_r<t_d$, as can happen in a large glitch, and a portion
$\Delta I_s$ of the core's moment of inertia participates in the
response, the fractional change in $\dot\nu$ is
\begin{equation}
\frac{\delta\dot\nu}{\vert\dot\nu\vert}=
-\frac{\Delta I_s}{I}
\frac{1}{1+({\rm e}^{t_d/t_r}-1){\rm e}^{-t/t_r}},
\end{equation}
depicted by the dashed curve in
Figure \ref{response_dresid}. Identification of this characteristic
non-exponential response through analyses of existing and future
timing data would give evidence that thermally-activated vortex slippage
plays a key role in post-glitch response. 

\item If a glitch is small enough that $t_d<<t_r$ is satisfied,
the glitch recovers as ${\rm e}^{-t/t_r}$. 

\end{enumerate}

Post-glitch response in the vortex slippage theory of this paper is
mathematically identical to the ``non-linear creep" regime of AAPS,
but with crucial quantitative differences.  In this paper, the scaling
of the recoupling time with the pinning energy $E_p$ is found to be
$t_r\propto E_p^{0.9}$ for the inner crust and $t_r\propto E_p^{3/5}$
for the outer core. By contrast, AAPS conclude that $t_r$ is {\em
independent} of $E_p$ in the ``non-linear" regime, and exponentially
dependent on $E_p$ in the ``linear regime".

To summarize, the vortex theory of AAPS and its extensions has two
regimes of vortex motion, the ``non-linear" and ``linear" regimes, while
the vortex slippage theory given here has only one, the ``non-linear"
regime. The vortex slippage theory presented here predicts a minimum
post-glitch response time of $t_d$, which is determined by the size of
the glitch. Vortex creep theory invokes a ``linear creep" regime that
is not realized for physically reasonable parameters, and that gives
very short recovery times without a waiting time $t_d$ before the
system recouples. By contrast, in
the vortex slippage theory developed in this paper, the total
post-glitch response time can only exceed $t_d$, giving strong
constraints on the theory. 

This paper presents the theory of thermally-activated vortex slippage,
with preliminary comparison with data; detailed comparisons with data
and constraints of the theory will be given in a forthcoming
publication. 
In Section \ref{overview}, the implications of post-glitch response are
described in terms of a simple but quite general model of the coupling
between the crust and the liquid interior. In Section \ref{dragregime}, the
response of a neutron star without pinning is reviewed. In Section
\ref{pinning}, pinning of vortices to inner-crust nuclei and
outer-core flux tubes is described. 
In Section \ref{slippage}, the theory of thermally-activated vortex
slippage is developed. In Section \ref{AE}, the mechanics and energetics of
vortex pinning is discussed. In Section \ref{relaxation}, 
the relaxation dynamics are calculated and
described. In Section \ref{timescales}, the coupling
timescales are estimated for the inner crust and the outer core. In Section
\ref{data}, preliminary comparisons of the calculations with observed
post-glitch response are made. In Section
\ref{comparison}, the results are compared with the previous
work of AAPS and its extensions. Appendix \ref{transition} determines
the conditions under which vortex pinning disappears, and vortex
motion enters the drag regime. Appendix
\ref{antiparallel} shows that vortex motion against the Magnus
force is strongly suppressed. Appendix \ref{random_pinning} shows that
vortex pinning is unlikely to occur in the denser regions of the inner
crust, including the pasta region, so that the ``superweak pinning''
regime proposed by AAPS, a regime that plays a key role in vortex
creep theory, is probably ruled out. 

\section{Post-glitch Response: General Considerations}

\label{overview}

\begin{figure*}[t]
\centering
\includegraphics[width=.6\linewidth]{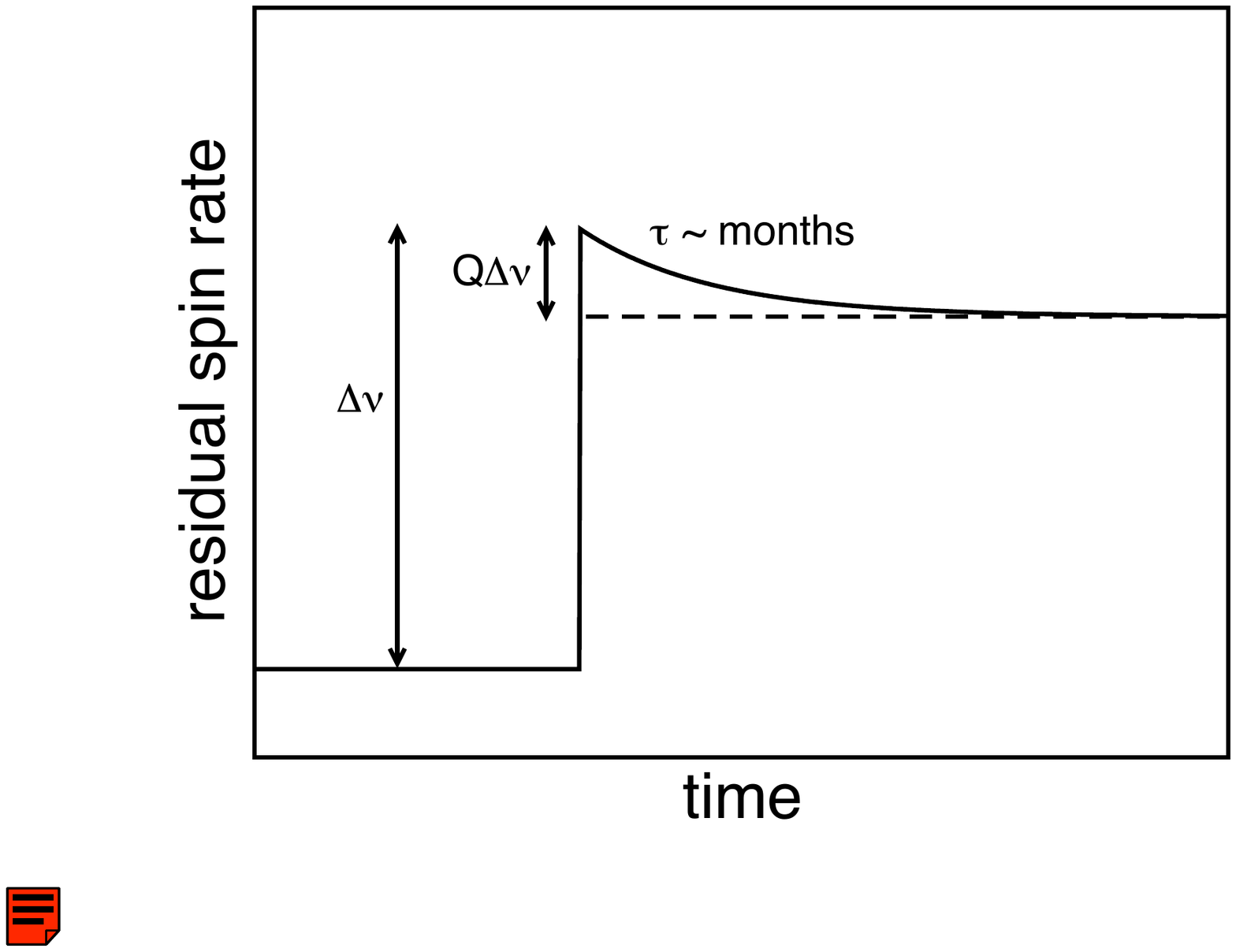} 
\caption{Schematic of a typical glitch, showing recovery by a fraction
$Q$.}
\label{typical_glitch}
\end{figure*}

Post-glitch response shows bewildering variety, with no clear
systematics \citep{lss00,wang_etal00}. Most glitches are consistent
with simple steps in the spin rate, with a recovery fraction $Q<<1$
over timescales of hundreds to thousands of days, as shown
schematically in Figure \ref{typical_glitch}. In many pulsars, the
Vela pulsar for example, the spin-down rate of the star does not
recover to its post-glitch value before the next glitch occurs
\citep{cordes_etal88,lyne_etal96,dodson_etal02,dodson_etal07}, 
indicating that rotational equilibrium is never reached; the internal
torque on the star never becomes constant. The post-glitch response of
such pulsars is problematic to study theoretically, since the glitch
cannot be described as a perturbation about a state of rotational
equilibrium. 

Consider the stellar crust plus charges, component $C$ of moment of
inertia $I_c$, coupled to the superfluid interior, component $S$ of
moment of inertia $I_s$ comprising the neutron superfluids of both the
inner crust and the core. Let a glitch occur at $t=0$. To account for
the fact that most glitches do not recover completely ($Q<1$), I
introduce a third component, component $L$, which imparts angular
momentum $J_l$ at the time of the glitch. Unlike $C$ and $S$, this
``loose screw'' is never in rotational equilibrium; it couples to $C$
only at the time of the glitch, but remains otherwise completely
decoupled. As described in Section
\ref{timescales}, much of the neutron star interior can remain decoupled
by glitches if there is pinning; decoupled components in either the
crust or the core could represent the loose screw that drives
glitches.

Assuming axisymmetry for simplicity, the rotational dynamics of the
system is given by 
\begin{equation}
I_c\dot{\Omega}_c+\int dI_s\, \dot{\Omega}_s=N_{\rm ext}+J_l\,\delta(t), 
\label{jcons}
\end{equation}
where $N_{\rm ext}$ is the external spin-down torque on
$C$. Sufficiently long after the glitch, the system will relax to {\em
spin-down equilibrium}, with
$\dot{\Omega}_c=\dot{\Omega}_s=\dot{\Omega}_0$, where $\dot{\Omega}_0$
is the equilibrium spin-down rate of both $C$ and $S$. The external
torque is given by $N_{\rm ext}=I\dot{\Omega}_0$, where $I\equiv
I_c+I_s$. (Note that $I$ excludes component $L$). Defining the {\em
lag} $\omega(r,t)\equiv\Omega_s(r,t)-\Omega_c(t)$, where $r$ is the
polar radius, Equation [\ref{jcons}] can be rewritten as
\begin{equation}
\dot{\Omega}_c=-\frac{1}{I}\int dI_s\, \dot{\omega}(r,t)+\dot{\Omega}_0
+\frac{J_l}{I}\delta(t).
\label{z7}
\end{equation}
The change in spin rate across the glitch is 
\begin{equation}
\Delta\Omega_c\equiv\Omega_c(0+)-\Omega_c(0-)=-\frac{1}{I}\int dI_s\,
\left[\omega(r,0+)-\omega(r,0-)\right]
+\frac{J_l}{I},
\label{one}
\end{equation}
and the solution to Equation [\ref{z7}] for $t>0$ is 
\begin{equation}
\Omega_c(t)-\Omega_c(0+)=
-\frac{1}{I}\int dI_s\,\left[\omega(r,t)-\omega(r,0+)\right]+\dot{\Omega}_0\,t, 
\label{two}
\end{equation}
assuming constant $N_{\rm ext}$. The {\em spin rate residual} is the
difference between the post-glitch spin rate and the value it would
have had if the glitch had not occurred:
\begin{equation}
\delta\Omega_c(t)\equiv \Omega_c(t)
-\left[\Omega_c(0-)+\dot{\Omega}_0t\right].
\label{residdef}
\end{equation}
Adding Equations [\ref{one}] and [\ref{two}], and using the definition of
the spin residual, gives
\begin{equation}
\delta\Omega_c(t)=-\frac{1}{I}\int dI_s\, \left[\omega(r,t)-\omega(r,0-)\right]
+\frac{J_l}{I}.
\label{z3}
\end{equation}
Of general interest is coupling between $C$ and $S$ that depends only on
the local lag $\omega(r,t)$:
\begin{equation}
\dot{\Omega}_s(r,t)=f(\omega(r,t)), 
\label{coupling}
\end{equation}
where the coupling could be linear or non-linear in
$\omega$. The state of spin-down equilibrium, $\omega_0(r)$, is given
by the solution to 
\begin{equation}
\dot{\Omega}_0=f(\omega_0(r)).
\end{equation}
Let $S$ and $C$ be in spin-down equilibrium just before the
glitch. Then 
\begin{equation}
\delta\Omega_c(t)=-\frac{1}{I}\int dI_s\, \left[\omega(r,t)-\omega_0(r)\right]
+\frac{J_l}{I}.
\label{resid}
\end{equation}
At late times after the glitch, $\omega(r,t)$ recovers to
$\omega_0(r)$, and the residual becomes
\begin{equation}
\lim_{t\rightarrow\infty}\delta\Omega_c(t)=\frac{J_l}{I}=(1-Q)\Delta\Omega_c.
\label{Jl}
\end{equation}
From Equation [\ref{jcons}], $I_c\Delta\Omega_c=J_l$, so that
\begin{equation}
Q=\frac{I_s}{I}.
\end{equation}
That many observed glitches have $Q<<1$ with recovery of
the spin-down rate to its pre-glitch value suggests, in the context of
this simple model, that glitches are driven by an out-of-equilibrium
component, and that only a small fraction of the superfluid is
associated with post-glitch response.\footnote{
More generally, we might expect that some fraction $f$ of the angular momentum
imparted to $C$ at the time of the glitch comes from $L$, with the
remainder coming from $S$. For this situation, $Q$ lies in the range
\begin{equation}
\frac{I_s}{I}\le Q\le 1.
\label{Qrange}
\end{equation}
where $Q=I_s/I$ corresponds to $f=1$, while $Q=1$ corresponds to
$f=0$. Henceforth I take $f=1$, a restriction that does not affect the
conclusions.}

\section{Dynamical Response without Vortex Pinning}

\label{dragregime}

The dynamics of a superfluid is determined by the motion of the
vortices, which is in turn determined by the dissipative forces
on the rotating superfluid. In regions of the star where there is no
vortex pinning, vortex motion is determined by drag forces. Here I
review the dynamics in this regime. 

Suppose a straight vortex is moving at velocity $\vbf_v$ with respect
to its environment. The environment could be the nuclear lattice of
the inner crust, or the proton-electron fluid of the outer core. 
The  motion of a straight vortex is given by balance between
the Magnus force on the vortex and the drag force:
\begin{equation}
\rho_s\kappabf\times\left(\vbf_v-\vsbf\right)-\eta \vbf_v = 0, 
\label{drag}
\end{equation}
where $\vsbf$ is the flow velocity of the ambient superfluid in the
rest frame of the environment, $\rho_s$
is the superfluid mass density, $\kappa=h/2m_n$ is the vorticity
quantum where $m_n$ is the neutron mass$, \kappabf$ is directed along
the vortex, and $\eta$ is the drag coefficient.\footnote{
In the low-temperature environment of a neutron star, a vortex has negligible
effective mass; its velocity is determined entirely by $\vsbf$ and
local forces \citep{bc83}.}  In cylindrical
coordinates $(r,\phi,z)$, with $z$ and $\kappabf$ along the
rotation axis, the vortex velocity is given by the solution of
Equation [\ref{drag}]; see \citet{be89} and \citet{eb92}:
\begin{equation}
\vbf_v=v_s\left(\frac{1}{2}\sin
2\theta_d\,\hat{r}+\cos^2\theta_d\,\hat{\phi}\right)=
v_s\cos\theta_d\left(\sin\theta_d\,\hat{r}+\cos\theta_d\,\hat{\phi}\right)
=r\omega\cos\theta_d\,\hat{n},
\label{vv}
\end{equation}
where the {\em dissipation angle} $\theta_d$ is given by 
\begin{equation}
\tan\theta_d\equiv\frac{\eta}{\rho_s\kappa}.
\label{theta}
\end{equation}
The vortex moves in direction $\hat{n}$, at an angle $\theta_d$ with
respect to $\vsbf$ with a component away from the rotation axis for
positive lag $\omega$. An important feature of vortex dynamics is
that local forces determine the vortex velocity, not the vortex acceleration.

The dissipation angle is simply related to the relaxation time of the
system. Assuming axisymmetry, the angular acceleration of the
superfluid follows from vorticity conservation (AAPS; \citealt{leb93})
\begin{equation}
\dot{\Omega}_s(r,t)=-\frac{1}{r}\left(2\Omega_s(r,t)+
r\frac{\partial}{\partial r}\Omega_s(r,t)\right)
\vbf_v\cdot\hat{r}, 
\label{sfeom1}
\end{equation}
and is determined by the radial component of the vortex velocity which
vanishes in the limit of zero dissipation ($\theta_d=0$).  
Assuming uniform superfluid rotation $\partial\Omega_s/\partial
r=0$ for simplicity, the equation of motion for the angular velocity lag
$\omega\equiv\Omega_s-\Omega_c$ is, from Equation [\ref{jcons}], 
\begin{equation}
\dot{\omega}=\frac{I}{I_c}(\vert\dot{\Omega}_0\vert+\dot{\Omega}_s)
- \frac{J_l}{I_c}\delta(t).
\label{lagdot}
\end{equation}
Linearizing Equation [\ref{sfeom1}], assuming
constant $\theta_d$, and using Equation [\ref{vv}], gives
\begin{equation}
\dot{\Omega}_s=-\omega\,\Omega_s\sin 2\theta_d, 
\end{equation}
where $\Omega_s$ is the unperturbed rotational velocity of the
superfluid. 
Assuming spin-down equilibrium with $\omega(0-)=\omega_0$ just before
the glitch, the solution is 
\begin{equation}
\omega(t)=\omega_0+(\omega(0+)-\omega_0){\rm e}^{-t/t_r}
\qquad
t_r=\frac{I_c}{I}\frac{1}{\Omega_s\sin 2\theta_d}.
\label{tr_drag}
\end{equation}
From Equation [\ref{lagdot}], the equilibrium lag is 
\begin{equation}
\omega_0=\frac{I}{I_c}t_r\vert\dot{\Omega}_0\vert.
\end{equation}

In the $C+S+L$ model of Section \ref{overview}, with a spin glitch driven by
angular momentum from components $L$, the response for 
uniform rotation follows from Equations [\ref{resid}], [\ref{Jl}], and
[\ref{jcons}]: 
\begin{equation}
\frac{\delta\Omega_c(t)}{\Delta\Omega_c}=1-Q(1-{\rm e}^{-t/t_r})
\qquad
Q=\frac{I_s}{I}.
\label{friction}
\end{equation}
In this linear treatment, the timescale of relaxation is independent
of the size of the glitch. 

As described below, vortices are expected to pin to nuclei in the
inner crust and to flux tubes in the outer core (if the outer core is
a type II superconductor). Pinning, though, is not expected to occur
everywhere in the star. Let us therefore first consider calculated
values for the drag strength without pinning, beginning with the inner
crust. 

If vortices in the inner crust move in the drag regime, the dominant drag
at low lag arises from scattering of the moving vortex against the
electron-phonon system of the lattice. \citet{jones90a} finds $3\times
10^{-5}\lap \sin
2\theta_d\lap 10^{-6}$, with a typical relaxation time of 
\begin{equation}
t_r=0.2\,\frac{I}{I_c}\left(\frac{\nu}{10\mbox{ Hz}}\right)^{-1}
\left(\frac{\sin 2\theta_d}{10^{-6}}\right)^{-1}\mbox{ days}, 
\end{equation}
where $\nu$ is the stellar spin rate. 
Relaxation through this process is shorter than observed, but the
timescale increases significantly as the interaction energy between a
nucleus and a vortex is reduced. 

\citet{be89} studied the scattering of electrons with the electron
cloud induced around a vortex through its interaction with nuclei. They
found $\sin 2\theta_d\lap 10^{-10}$, corresponding to a relaxation
time of 2000 d for $\nu=10$ Hz. 

\citet{feibelman71} considered the scattering of electrons with the
magnetic moments of thermally-excited neutrons in the vortex core. The
results are strongly sensitive to the neutron pairing gap and to the
temperature. From these calculations, using a range of plausible
pairing gaps, $\sin 2\theta_d$ can range from $\sim 10^{-8}$ to
effectively zero. 

In the outer core, entrainment of the neutron and proton mass currents
when both species are superfluid endows a vortex with a magnetic field
of $B_v\sim 10^{14}$ G \citep{als84}.  Electron scattering with the
strongly-magnetized vortex cores gives $\sin 2\theta_d \sim 10^{-3}$
\citep{as88}. 
Hence, if there is no pinning in the core, electron scattering is so
effective that the neutral and charged fluids of the core corotate
over timescales $\gap 10^3\,\Omega_c^{-1}$. The core liquid is thus 
typically considered
to corotate with the crust. \citet{ss95} and \citet{ssct95} have obtained the
opposite conclusion; 
ignoring the natal magnetic field of the neutron star, 
\citet{ss95} and \citet{ssct95} argue that 
magnetic flux tubes of the outer core form clusters around
vortices, which could increase the coupling time to observable
timescales of days or longer. They suggest that post-glitch response
has a core component. The effects of the natal magnetic field on these
conclusions have not been studied, and could prove to be important.

The point of
view taken in this paper is that post-glitch response could 
involve both the inner-crust and core superfluids, and both
possibilities will be considered. 

\section{Vortex Pinning}

\label{pinning}

\subsection{Pinning of Vortices to Nuclei in the Inner Crust}

As originally pointed out by \citet{pines71} and elaborated upon by
\citet{ai75} and \citet{alpar77}, vortices of the inner crust interact
with nuclei. The main contribution to this interaction is the density
dependence of the superfluid condensation energy per particle. When
the increase of energy associated with the destruction of
superfluidity within the vortex core exceeds the corresponding
increase inside the nucleus, the energy cost is minimized if the
vortex overlaps with the nucleus, giving an attractive interaction. If
the inequality goes the other way, the energy is minimized if the
vortices can get as far from nuclei as possible, in which case
vortices could pin to the interstices of the nuclear lattice.  Pinning
is predicted to be strongest in the denser regions of the inner
crust. Recent pinning calculations generally agree on the magnitude
and density dependence of $E_p$, but not its sign.
\citet{dp06} find that the vortex-nucleus interaction
energy is $E_p\simeq 3$ MeV, and attractive, for baryon densities
$\rho_b$ in the narrow range $3\times 10^{13}<\rho_b<5\times 10^{13}$
\gcc. \citet{abbv07} find a similar interaction strength in this
density range, though repulsive. Both calculations show a sharp drop
in $\vert E_p\vert$ to nearly zero above $\rho_b\simeq 6\times
10^{13}$ \gcc. These calculations did not address the spatial
dependence of the interaction, and so only estimates of the pinning
force based on the relevant length scales could be obtained. The
length scale of the interaction is of order the coherence length
$\sim\xi_n$ of the neutron superfluid, and so the force per pinning
site is $F_p\sim E_p/\xi_n$ for pinning to nuclei. This estimate
should apply for a repulsive interaction as well. A vortex is an
object with finite rigidity (see below), and in order to move through
the lattice it must be brought close to nuclei that repel the vortex
over a length scale $\xi_n$. Hence, nuclear and interstitial pinning
should be very similar.\footnote{For a repulsive interaction,
\citet{dp06} estimate the interaction force per site to be $F_p\sim
E_p/a$, where $a$ is the nuclear spacing. Since $a$ is much larger
than $\xi_n$ in regions where they find the interaction to be
repulsive, they conclude that interstitial pinning is relatively
unimportant, contrary to the argument given here.} \citet{eb88}, using
Ginzburg-Landau theory, obtained $E_p\simeq 10$ MeV; Ginzburg-Landau
theory is however inadequate for this problem, since the superfluid coherence
length is comparable to the size of a nucleus.

Above a baryon density $\rho_b \simeq 6\times 10^{13}$ \gcc, the pinning
situation is unclear. The coherence length exceeds
the nuclear spacing \citep{schwenk_etal03,cao_etal06}, which weakens
pinning. At $\rho_b\simeq 10^{14}$ \gcc, the nuclei acquire
non-spherical pasta shapes
\citep{rpw83}. Pinning to nuclear pasta is probably weaker than
pinning to spherical nuclei. Hence, the base of the inner crust could
represent a physically distinct region with relatively weak pinning or
no pinning. Pinning to nuclear pasta has been studied by 
\citet{mochizuki1995self} and
\cite{mochizuki1997exotic}, who suggest that a vortex can induce
nuclear rod formation along its length, becoming self-pinned. Pinning
without idealized geometries has not been studied; a first attempt at
this problem is given in Appendix \ref{random_pinning}.

\begin{figure*}[t]
\centering
\includegraphics[width=.6\linewidth,angle=-90]{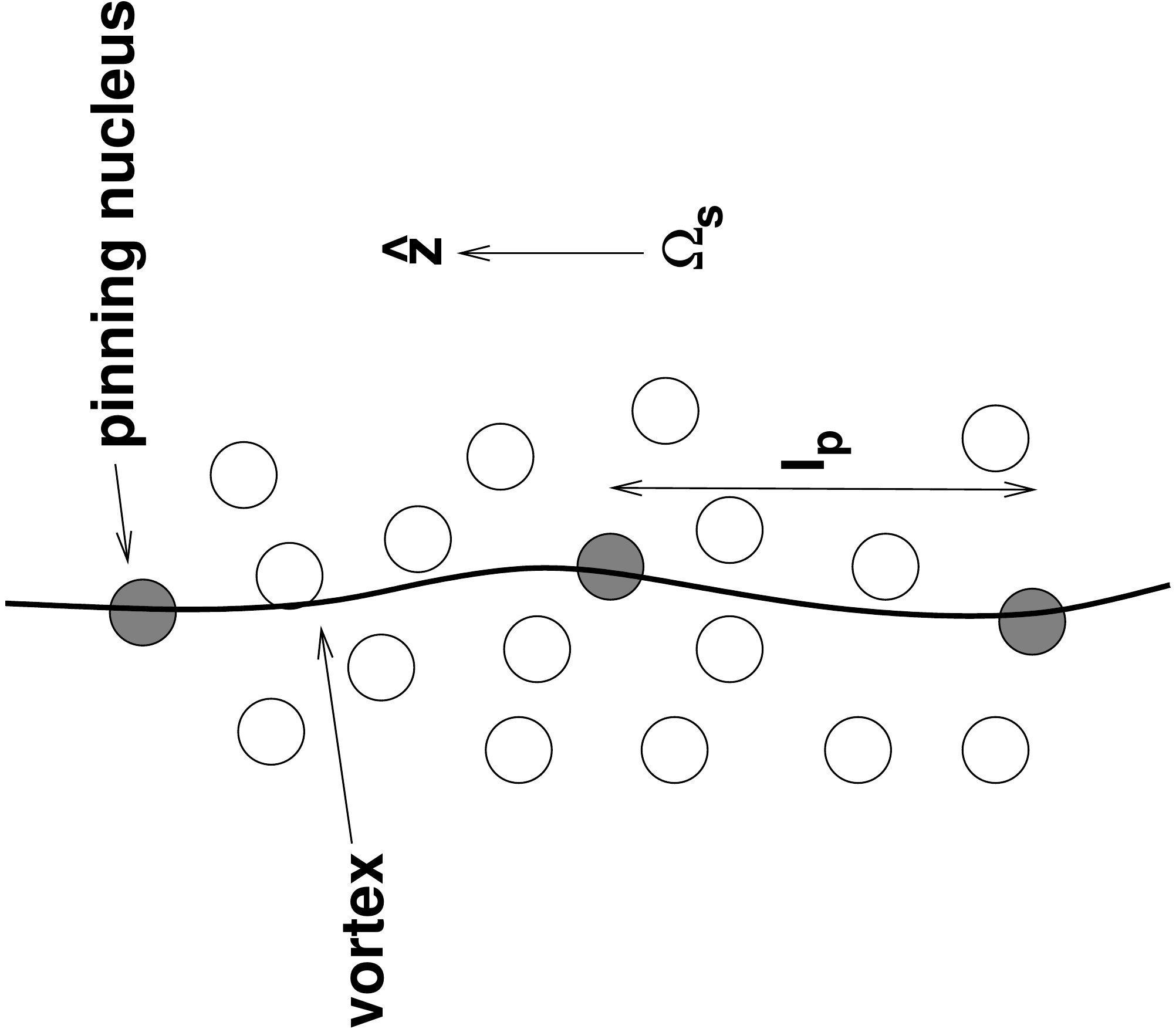}
\caption{Pinning of a vortex to nuclei. The vortex generally follows
the rotation axis, but because it has finite rigidity, it cannot bend to
intersect every nucleus; rather, the vortex will pin to the shaded
``pinning nuclei'', generally missing the unshaded nuclei. The
relative importance of the pinning force and the tension force
determines the pinning length $l_p$. The pinning situation is
illustrated here for an amorphous lattice, but these considerations
are essentially unaltered for pinning in a regular lattice, since the
crystal planes will not generally be aligned with the rotation vector
of the superfluid. }
\label{nuclear_pinning}
\end{figure*}

If a vortex could bend to intersect nuclei separated by $a$, the
pinning force per unit length would be 
\begin{equation}
f_p= \frac{F_p}{a}\simeq\frac{E_p}{a\xi_n}, 
\end{equation}
typically of order $10^{17}$ dyn cm$^{-1}$ in the inner crust. This
force is reduced by the vortex's large self-energy, or {\em tension},
associated with the kinetic energy of the flow about the vortex
\citep{thomson1880,fetter67}: 
\begin{equation}
T_v=\frac{1}{2}\int d^2r\, \rho_s \left(\frac{\kappa}{r}\right)^2
=\frac{\rho_s\kappa^2}{4\pi}\Lambda 
\qquad \qquad
\Lambda\equiv\ln\frac{l_v}{\xi_n}, 
\label{tension}
\end{equation}
where $\rho_s$ is the density of free superfluid neutrons.  A lower
cut-off at the vortex core radius and an upper cut-off at the
inter-vortex separation $l_v$ have been introduced; the logarithmic
factor $\Lambda$ is typically $\simeq 3$, as assumed
henceforth. Because the vortex is stiff, it cannot bend to intersect
every nucleus, but bends over a length scale $l_p>a$, as shown in
Figure \ref{nuclear_pinning}.

Throughout much of the inner crust, most of the neutron
superfluid is non-dissipatively {\em entrained} by the nuclei and does
not participate in the superfluid flow. \citet{chamel05,chamel12}
finds that $\sim 90$\% of the neutron mass is entrained in the denser
regions of the inner crust. The effects of entrainment can be treated
by taking
\begin{equation}
\rho_s=f_c\rho_n
\end{equation}
where $\rho_n$ is the total mass density in neutrons, both bound to
nuclei and unbound, $\rho_s$ is the density of superfluid {\em
conduction neutrons} that are not entrained by nuclei, and $f_c\sim
0.1$ is the fraction of of the neutron fluid comprising superfluid
conduction neutrons. Accounting for nuclear entrainment, the typical
vortex self energy is
\begin{equation} T_v\simeq 0.6\, f_c\,
\rho_{n,14}\mbox{ MeV fm$^{-1}$},
\label{tension1}
\end{equation}
where $\rho_{n,14}$ is the total superfluid mass density in units of
$10^{14}$ \gcc. Tension makes the vortex difficult to bend over the typical
nuclear spacing $a\simeq 50$ fm of the inner crust.

Dynamical simulations of a vortex in a random potential at zero
temperature by \citet{link09}, accounting for vortex tension, show
that pinning is inevitable below a critical value of the superfluid
flow speed with respect to the lattice, but that the pinning force per
unit length $f_p$ is reduced by a factor $a/l_p$, where $l_p$ is the
characteristic bending length of a vortex, typically $10-30a$ (if
nuclear entrainment is neglected). A variational estimate
(\citealt{link12a}; see also \citealt{lc02}) gives for the pinning
force per unit length
\begin{equation}
f_p=\frac{F_p}{l_p}\simeq 
\frac{E_p}{a\xi_n}\left(\frac{2E_p}{3aT_v}\right)^{1/2} \qquad
\mbox{where}
\qquad
\frac{l_p}{a}=\left(\frac{3aT_v}{2E_p}\right)^{1/2},
\label{lp}
\end{equation}
typically of order $10^{16}$ dyn cm$^{-1}$. A similar number was
obtained by \citet{gp12}. 

The critical velocity difference $v_s$ in the rest frame of the
solid above which pinning becomes unstable follows by equating the
Magnus force per unit length to the pinning force per unit
length. In terms of the critical lag $\omega_{\rm crit}=v_{s,{\rm
crit}}/r$ at polar radius $r$, accounting for entrainment by nuclei,
the Magnus force is 
\begin{equation}
f_c\rho_n\kappa r\omega_{\rm crit} =f_p. 
\end{equation}
Combining with Equations [\ref{tension1}] and [\ref{lp}] gives
\begin{equation}
\omega_{\rm crit}=\frac{E_p}{rf_c\rho_s\kappa\xi_n l_p}\sim 
4\,
\rho_{n,14}^{-3/2}\,E_p(\mbox{MeV})^{3/2}\left(\frac{f_c}{0.1}\right)^{-3/2}
\left(\frac{\xi_n}{10\mbox{ fm}}\right)^{-1}
\left(\frac{r}{\mbox{10 km}}\right)^{-1}
\,
\mbox{ \rads}.
\label{omc_crust}
\end{equation}
Note that entrainment of the neutron superfluid
by nuclei increases $\omega_{\rm crit}$ by reducing both the tension
and the Magnus force. 

For small $E_p$, thermal motion of the vortex precludes pinning. As
shown in Appendix \ref{transition}, pinning disappears for 
\begin{equation}
E_p<0.04\mbox{ MeV}\,
\left(\frac{kT}{\mbox{10 keV}}\right)^{3/2}
\left(\frac{a}{\mbox{50 fm}}\right)
\left(\frac{\rho_{n,14}}{0.5}\right)^{1/3}
\left(\frac{f_c}{0.1}\right)^{1/3}.
\label{Epbound}
\end{equation}
Without entrainment, pinning vanishes for $E_p\lap 0.1$ MeV. 

As discussed in Appendix \ref{random_pinning}, pinning to nuclei is
likely to disappear above a baryon density of $\rho_b\simeq
6\times 10^{13}$. Above this density, the coherence length $\xi_n$
quickly begins to exceed the size of the Wigner-Seitz cell
\citep{schwenk_etal03,cao_etal06}, and vortices interact
primarily with spatial variations in the number density of
nuclei. The effective pinning energy becomes so low that thermal
fluctuations preclude pinning unless the interaction energy per
nucleus $E_p$ exceeds $\simeq 0.4$ MeV.
Recent pinning calculations show a sharp drop in $E_p$ to nearly zero
above $\rho_b\simeq 6\times 10^{13}$ \gcc\ \citep{dp06,abbv07}.  These
effects appear to eliminate ``superweak pinning'' proposed
by AAPS to exist in this density regime; typical values of $E_p$ for
``superweak pinning'' used in vortex creep theory are $\simeq 0.3$ MeV (\eg,
\citealt{alpar_etal89}). For $\rho_b\gap 6\times 10^{13}$ \gcc\ in the
inner crust, pinning probably does not occur at all, and vortex motion
enters the drag regime.

\subsection{Pinning of Vortices to Flux Tubes in the Outer Core} 

The protons of the outer core are predicted to form a type II
superconductor. As the protons condensed when
the star was young, the very high electrical conductivity of the
relativistically-degenerate electrons prevented Meissner expulsion of
the core's natal magnetic field
\citep{bpp69a}, and the field formed a dense tangle of magnetic flux
tubes with which vortices interact. This primarily magnetic
interaction pins the vortices to the magnetic tangle which is frozen
to the superconducting fluid. The flux tube tangle will
be treated as completely immobile under the stresses exerted by the
vortex array. 

Overlap of a vortex line and
a flux tube is energetically favored because the volume of uncondensed
fluid is minimized by such overlap; the interaction energy is $\sim
0.1$ MeV per junction \citep{srinivasan_etal90}. A far larger
contribution to the interaction energy is the magnetic interaction
between the two structures. The magnetic field in a flux
tube is $B_\Phi\sim 10^{15}$ G \citep{als84}. The magnetic interaction
energy between a vortex and a flux tube $E_p$ is of order $B_v B_\Phi
V$, where $V$ is the overlap volume, and has been estimated to be
$\sim 100$ MeV by a number of authors 
\citep{jones91,mendell91a,chau_etal92,rzc98,link12b}. 
The angle-averaged interaction energy
for the intersection of a vortex with a flux tube is \citep{link12b}
\begin{equation}
E_p \simeq 10^2\, 
\left(\frac{m_p^*/m_p}{0.5}\right)^{-1/2}
\left(\frac{\vert\delta m_p^*\vert/m_p}{0.5}\right)
\left(\frac{x_p}{0.05}\right)^{1/2} 
\left(\frac{\rho_{n,14}}{4}\right)^{1/2}
\mbox{ MeV}, 
\label{Ep}
\end{equation}
where $m_p$ is the bare proton mass, $m_p^*\equiv m_p+\delta m_p^*\sim
m_p/2$ \citep{sjoberg,ch06} is
its effective mass, and $x_p\simeq 0.05$ is the proton mass fraction. 
The range
of the interaction between a vortex and flux tubes is of order the
characteristic radius of a flux tube, or the London length, given by
\citep{als84},
\begin{equation}
\Lambda_*=50\, 
\left(\frac{m_p^*/m_p}{0.5}\right)^{1/2}
\left(\frac{x_p}{0.05}\right)^{-1/2}
\left(\frac{\rho_{n,14}}{4}\right)^{-1/2}\mbox{ fm.}
\label{Lambda}
\end{equation}
The pinning force is $F_p\sim E_p/\Lambda_*$, about 1 MeV
fm$^{-1}$. For a single vortex immersed in a tangle of flux tubes, the
average length between intersections will equal the average distance
between flux tubes \citep{link12b}, 
\begin{equation}
l_p\sim l_\Phi\simeq 5\times 10^3\, B_{12}^{-1/2}\mbox{ fm},
\label{lphi}
\end{equation}
where $B_{12}$ is the magnetic field in units of $10^{12}$ G. Note
that the relevant field is the average {\em internal} field of the
star, which can be significantly larger than the dipolar component 
\citep{braithwaite09}. Because a vortex must cross a ``fence'' of flux
tubes in order to move, the pinning spacing is not increased by the
vortex self energy, as is the case for pinning to nuclei in the inner
crust.

In the outer core, entrainment between the protons and neutrons,
though essential for pinning, has only a small effect on the free
neutron density \citep{ch06,link12b}; $f_c$ is effectively unity in evaluating both the
tension and the Magnus force. 
The critical angular velocity difference $\omega_{\rm crit}$ between
the neutron superfluid and charged component to which the flux tubes
are frozen, is given by 
\begin{equation}
\rho_n\kappa r\omega_{\rm crit} l_\Phi=\frac{E_p}{\Lambda_*}, 
\label{critical}
\end{equation}
giving \citep{link12b}
\begin{equation}
\omega_{\rm crit} \sim 10^{-1}\, 
\left(\frac{x_p}{0.05}\right)
\frac{\vert\delta m_p^*\vert}{m_p^*}\left(\frac{E_p}{100\mbox{
MeV}}\right)
\left(\frac{r}{\mbox{10 km}}\right)^{-1}
\,B_{12}^{1/2}\mbox{ \rads}. 
\label{omc_core}
\end{equation}

As shown in Appendix \ref{transition}, thermal motion 
eliminates pinning when 
\begin{equation}
E_p<0.3\mbox{ MeV}\,
\left(\frac{kT}{\mbox{10 keV}}\right)^{3/2}B_{12}^{-1/2}. 
\end{equation}
The large value of $E_p$ of Equation [\ref{Ep}] suggests 
that pinning occurs wherever type II protons coexist with
superfluid neutrons. 

\subsection{Dissipation with Vortex Pinning}

\label{dissipation_with_pinning}

Pinning causes differential rotation between 
superfluid and the crust to develop as the crust is spun down by external
torque. If a vortex segment unpins, it will be dragged though its
environment at relatively high velocity, and the dissipation will
be generally stronger than the processes discussed in Section
\ref{dragregime} for small velocities. 
In the inner crust, pinning of vortices to nuclei sustains
velocity differences of $\sim r\omega_{\rm crit}\sim 10^6$ \cms. The
dominant contribution to vortex drag on an unpinned segment is the
excitation of Kelvin phonons on the vortex
\citep{eb92,jones92}, giving $\sin 2\theta_d$ as large as $\sim 0.1$. 
For $\omega\lap 10^{-2}$ \rads, Kelvin phonon production is expected
to be greatly reduced \citep{jones92}, with a corresponding reduction
in the drag. The reduction has not been calculated.

In the outer core, pinning of vortices to flux tubes produces velocity
differences of $\sim r\omega_{\rm crit}\sim 10^5$ \cms. The strength
of vortex drag has not been calculated in this regime, but it is
reasonable to expect the dissipation to be stronger than the low
velocity limit considered by \citet{als84}, for which $\sin
2\theta_d\sim 10^{-3}$. The motion of vortex segments with
respect to flux tubes will also excite Kelvin phonons on the vortices,
and produce wave excitations on flux tubes; both processes will
contribute to the dissipation, and have not been
calculated. Dissipation due to the forcing of vortices through flux
tubes at super-critical velocities has been studied by
\citet{link03}.

In the estimates below, $\sin 2\theta_d>10^{-2}$ will be assumed for
the inner crust, and $\sin 2\theta_d>10^{-3}$ will be assumed for the
outer core. The coupling timescales calculated in Section \ref{timescales}
depend very weakly on these choices.

\section{Thermally-activated Vortex Slippage} 

\label{slippage}

As a result of vortex pinning to nuclei in the inner crust,
and to flux tubes in the outer core, the lag $\omega$
approaches its local critical value as the charged
component is spun down by the electromagnetic torque. 
Before $\omega_{\rm crit}$ is reached, though, vortices
will slip over their pinning barriers through hermal activation. I
now calculate the vortex slippage velocity. 

The superfluid circulation speed around a vortex is inversely
proportional to the distance from the vortex. The ratio of the pinning
interaction length to the pinning spacing is $\sim 0.2$ in the inner
crust, and $\sim 0.1$ in the outer core. A pinned vortex will
henceforth be treated as a thin string with tension in classical continuum
mechanics. 

\begin{figure*}[t]
\centering
\includegraphics[width=.6\linewidth,angle=-90]{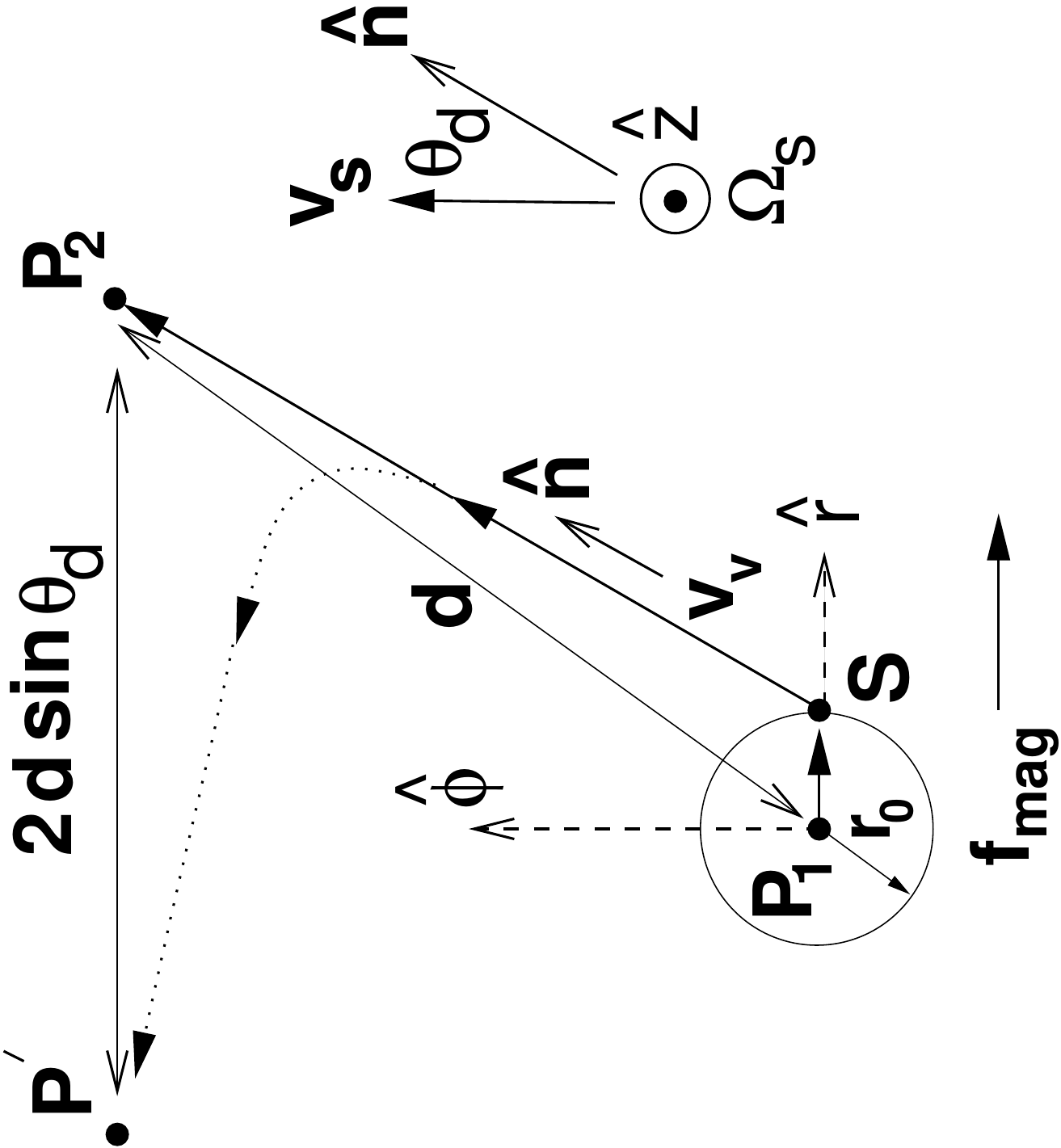}
\caption{Trajectories of vortex segments. The vortex and rotation
axes are out of the page. A vortex segment initially pinned along an
axis $P_1$ of pinning sites of spacing $l_p$ is thermally activated
under the Magnus force to saddle-point configuration $S$. The segment
then translates under dissipation in direction $\hat{n}$ at velocity
$\vbf_v$ before repinning at a new axis $P_2$ of pinning sites, a
distance $d$ from $P_1$. In this process, the segment moves a
distance $d\sin\theta_d$ away from the rotation axis. Competing
processes involving motion $d\sin\theta_d$ towards the rotation
axis, such as to axis $P^\prime$ along the dotted trajectory, 
are prevented by the strong vortex self
energy; see Appendix \ref{antiparallel}.} 
\label{trajectories}
\end{figure*}

\begin{figure*}[!ht]
\centering
\includegraphics[width=.9\linewidth,angle=-90]{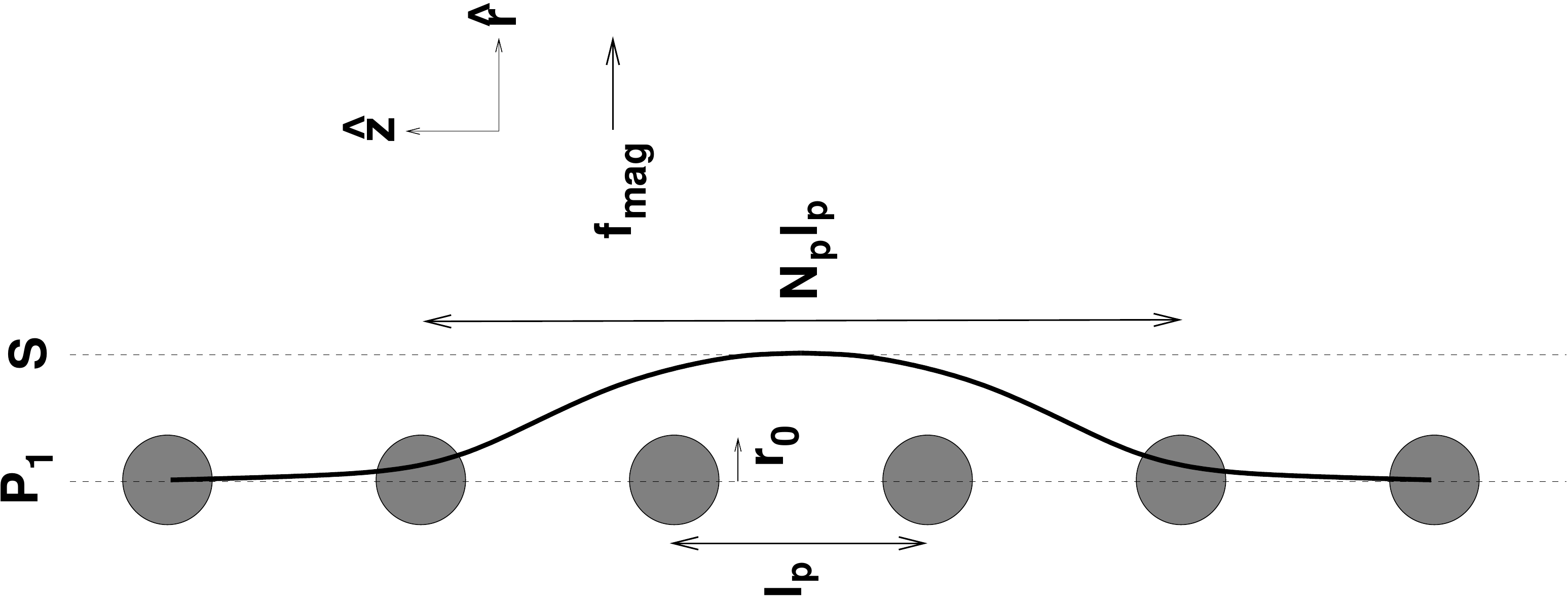} 
\caption{Saddle-point configuration in a simplified geometry. 
A vortex initially pinned along a
linear array of pinning sites, nuclei in the inner crust or
flux tube intersections in the outer core, is thermally activated to a
saddle-point configuration of length $N_pl_p$ along axis $S$. The
number of pinning bonds that must break is set by the relative
strengths of vortex tension and the pinning energy (through ${\cal T}$), and
is generally significantly larger than unity. For the inner crust with
nuclear spacing $a$, the
pinning spacing is $\gap 10a$, depending on the pinning strength. In the outer
core, the pinning spacing is of order the flux tube spacing $l_\Phi$.}
\label{saddlepoint}
\end{figure*}

\begin{figure*}[!ht]
\centering
\includegraphics[width=.6\linewidth,angle=-90]{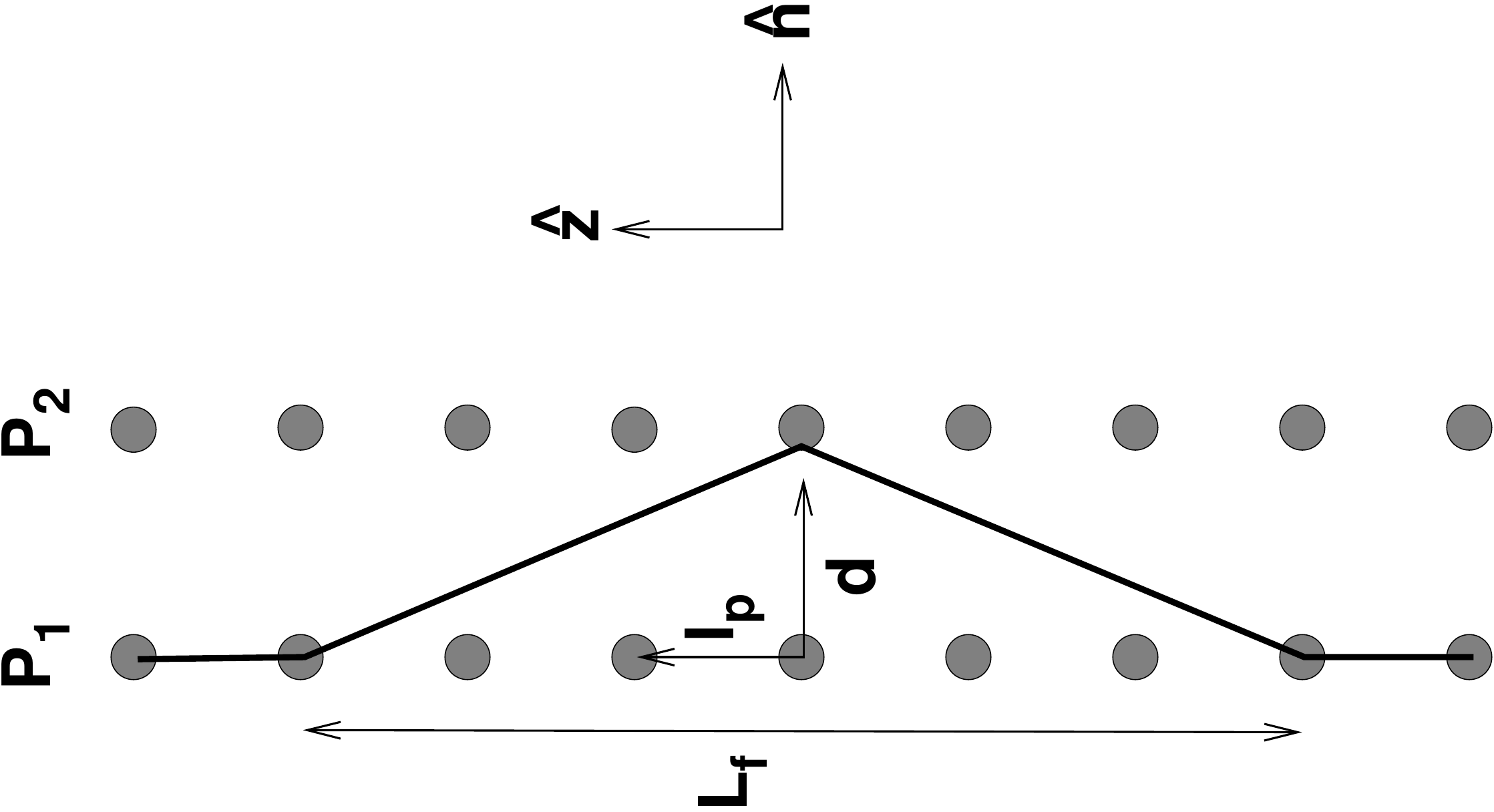}
\caption{New pinned state reached by a vortex segment in a
simplified geometry. For the
segment to reach a stable configurations as at axis $P_2$, it must
increase its length to a final length $L_f$, greater than the
pinning length $l_p$, through an ``unzipping'' process (see text). For
the inner crust, $d$ is approximately the nuclear spacing $a$. In the
outer core, $d$ is approximately the flux tube spacing $l_\Phi$.}
\label{repinned}
\end{figure*}

Consider pinning of a vortex along axis $P_1$; see
Figure \ref{trajectories}. Vortex motion occurs as follows. First,
thermal fluctuations excite a vortex segment of length $L_p$ to a
saddle-point configuration at $S$ that lies out of the range $r_0$ of the
pinning potential. The saddle-point configuration is shown in
Figure \ref{saddlepoint} for an idealized geometry of a linear array of
pinning centers spaced by $l_p$. The saddle-point configuration will
be described in detail below. In this unstable configuration, the
stress on the endpoints of the segment is increased by a factor
$L_p/l_p$, typically greater than unity. The segment will then
``unzip'' under the Magnus force until the vortex can reach another
prospective pinning axis a distance $d$ away; $d=a$ for the inner
crust and $l_\Phi$ for the outer core. During this stage, the
vortex segment is out of range of the pinning potential, and drifts at
velocity $\vbf_v$ in direction $\hat{n}$, given by Equation [\ref{vv}],
toward axis $P_2$. For pinning at $P_2$ to occur, the vortex
segment must unzip to length $L_f>L_p$, so that pinning to one site
along $P_2$ is stable. This configuration is shown in
Figure \ref{repinned}. The condition for mechanical equilibrium is
\begin{equation}
2T_v \sin\left[\tan^{-1}\left(\frac{2d}{L_f}\right)\right]\,\hat{n}-
\fbf_{\rm mag}L_f=\Fbf_p. 
\end{equation}
Typically $d<<L_f$. Using the same cylindrical coordinate system of Section
\ref{dragregime}, projecting onto $\hat{r}$ and $\hat{\phi}$, gives
\begin{equation}
\frac{4d}{L_f}T_v\sin\theta_d-f_{\rm mag}L_f=F_{p,r} \quad
\mbox{and}
\quad
\frac{4d}{L_f}T_v\cos\theta_d=F_{p,\phi}.
\end{equation}
Typically we are interested in $\theta_d<<1$. Noting that $F_{p,\phi}$
is limited by $F_p=R_p/r_0$, gives
\begin{equation}
L_f=\frac{4T_vdr_0}{E_p}. 
\label{Lf}
\end{equation}
As $E_p$ is lowered, $L_f$ typically exceeds $N_p$, and the vortex must
unzip from many pinning bonds to reach static equilibrium at $P_2$. 

Meanwhile, the same processes are occurring elsewhere along the
portions of the vortex that are still pinned along the initial pinning
axis $P_1$. The vortex slowly migrates, or {\em slips}, at angle
$\theta_d$ with respect to $\vbf_s$, but at a speed $v_v<<v_s$. As
shown in Appendix \ref{antiparallel}, competing transitions against
$\fbf_{\rm mag}$ to states such as $P^\prime$ are prevented by the
strong vortex tension unless the dissipation angle is unexpectedly
small. In the inner crust, $\sin 2\theta_d\gap 10^{-2}$ is expected.
With nuclear entrainment ($f_c=0.1$) and $E_p=0.04$ MeV, the lower
limit for pinning, the suppression factor is ${\rm e}^{-1.5}\simeq
0.22$, with much greater suppression at higher $E_p$. Transitions to
$P^\prime$ become important only for
\begin{equation}
\sin 2\theta_d< 1.5\times 10^{-4}\, 
E_p(\mbox{MeV})^{-1/2}\left
(\frac{f_c\rho_{n,14}}{0.5}\right)^{-1/2}
\left(\frac{a}{\mbox{50 fm}}\right)^{-1/2}
\left(\frac{kT}{\mbox{10 keV}}\right)
\qquad
\mbox{(inner crust)}
\end{equation}

In the outer core, where
$\sin 2\theta_d\gap 10^{-3}$ is expected, transitions to $P^\prime$ are
suppressed relative to transitions to $P_2$ by a
typical factor ${\rm e}^{-5\times 10^4}$, independent of $E_p$. Transitions to
$P^\prime$ become important only for 
\begin{equation}
\sin 2\theta_d<2\times 10^{-8}\, 
\left(\frac{\rho_{n,14}}{4}\right)^{-1}
\left(\frac{kT}{\mbox{10 keV}}\right)
B_{12}^{1/2}
\qquad
\mbox{(outer core)}. 
\end{equation}
Transitions against $\fbf_{\rm mag}$ can thus be ignored in both
the inner crust and the outer core. 

The vortex slippage process can be described in terms of a two-state
system; 1) the vortex is pinned, with zero velocity and zero energy,
or, 2) the vortex is unpinned, moving under drag at velocity
$\vbf_v$, with energy $A$ upon unpinning. The energy $A$ is the activation
energy for unpinning, specified below.  The partition function for
this two-state system is
\begin{equation}
Z=1+{\rm e}^{-\beta A}. 
\label{Z}
\end{equation}
Here $\beta\equiv(kT)^{-1}$ where $T$ is the temperature and $k$ is
Boltzmann's constant. For slow vortex slippage, ${\rm e}^{-\beta
A}<<1$ and $Z\simeq 1$. The slippage velocity of a typical vortex is
given by the statistical average
\begin{equation}
\langle\vbf_v\rangle_\beta = Z^{-1}{\rm e}^{-\beta A}\vbf_v\simeq 
r\omega\left(\frac{1}{2}\sin
2\theta_d\,\hat{r}+\cos^2\theta_d\,\hat{\phi}\right)
{\rm e}^{-\beta A}. 
\label{vslippage1}
\end{equation}
The terms that depend on $\theta_d$ refer to the forces exerted
on the unpinned vortex segment by the environment. In the inner crust,
the main dissipative force arises from the excitation of Kelvin
waves as a vortex segment moves past nuclei. In the outer core,
dissipation arises from electron scattering, but there will also be a
contribution from the interaction of the translating segment with
flux tubes; see Section \ref{dissipation_with_pinning}.

This treatment of vortex motion assumes the existence of a
well-defined pinned state, which requires $A>>kT$. The limit of no
pinning given by Equation [\ref{vv}] is recovered by taking $A\rightarrow
0$ and $Z\rightarrow 1$, the latter replacement corresponding to the
vortex having only one state of motion in this limit.

We are interested in the flow dynamics of the system over length
scales that exceed the inter-vortex spacing:
\begin{equation}
l_v=\left(\frac{\kappa}{2\Omega_s}\right)^{1/2}=3\times 10^{-3}
\left(\frac{\Omega_s}{100\mbox{ rad s$^{-1}$}}\right)^{1/2}\mbox{ cm}. 
\end{equation}
We obtain the desired limit by averaging the slippage velocity of
Equation [\ref{vslippage1}] for a single vortex over a bundle of $N>>1$
vortices. Consider a bundle of vortices enclosed by a contour that
spans an area $S$ that satisfies $\sqrt{S}>>l_v$, and average the
vortex slippage velocity over the bundle as follows:
\begin{equation}
\langle\vbf\rangle=\frac{1}{S}\int dS\, \langle{\vbf}\rangle_\beta.
\end{equation}
$S$ can be chosen so that it contains many vortices ($\sqrt{S}>>l_v$),
while being small enough that $A$, $\theta_d$, and $\omega$, which
vary over macroscopic scales, are nearly 
constant. This choice gives the same result as the velocity for a
typical vortex (Equation \ref{vslippage1}):
\begin{equation}
\langle\vbf\rangle=
r\omega\left(\frac{1}{2}\sin
2\theta_d\,\hat{r}+\cos^2\theta_d\,\hat{\phi}\right)
{\rm e}^{-\beta A}, 
\label{vslippage}
\end{equation}
Note that $\langle ...\rangle$ denotes both a statistical average and 
a spatial average. All vortices in
the bundle move in the same way on average, so
that $\theta_d$ after spatial averaging takes the same value that appears in
Equation [\ref{vslippage1}] for a single vortex.
The spatial averaging is similar to the
macroscopic averaging procedure introduced by \cite{bc83}. 

The equation of motion of the superfluid follows by replacing
$\vbf_v$ in Equation [\ref{sfeom1}] with $\langle\vbf_v\rangle$:
\begin{equation}
\dot{\Omega}_s(r,t)=-\frac{1}{r}\left(2\Omega_s(r,t)+
r\frac{\partial}{\partial r}\Omega_s(r,t)\right)
\langle\vbf_v\rangle\cdot\hat{r}.
\label{sfeom2}
\end{equation}

As an aside, I now connect the treatment so far to that of introducing
a {\em mutual friction} force into the fluid equations, 
usually taken to have the following form:
\begin{equation}
\fbf/\rho_s={\cal B}^\prime\,\omegabf_s\times(\vbf_s-\vbf_b)+
{\cal B}\,\hat\omega_s\times(\omega_s\times\left\{\vbf_s-\vbf_b\right\}), 
\end{equation}
where $\omegabf\equiv\nabla\times\vbf_s$ is the vorticity in the
inertial frame, $\vbf_b$ is the velocity of the background against
which vortices are dragged, and ${\cal B}^\prime$ and ${\cal B}$ are
the the mutual friction coefficients. 
In terms of the dissipation angle, the mutual friction coefficients are
\begin{equation}
1-{\cal B}^\prime=\cos^2\theta_d\,{\rm e}^{-\beta A}<<1 \qquad
{\cal B}=\frac{1}{2}\sin 2\theta_d {\rm e}^{-\beta A}<<1.
\end{equation}

Vortex pinning is sometimes considered to be equivalent to the limit of
high drag (\eg, \citealt{swc99,link06,gaj08a,vl08,ga09}). Assuming that
drag is the only force on a vortex, the vortex velocity in
Equation [\ref{vv}] goes to zero in the limit
$\eta\rightarrow\infty$. This description of vortex pinning is
inadequate because it ignores the forces exerted on the
vortex by the pinning lattice, which can immobilize the vortex even
for zero dissipation. As Equation [\ref{vslippage}] shows, the vortex speed
is $v_v={\rm e}^{-\beta A}\,v_s<<v_s$ for $\eta=0$
($\theta_d=0)$. Dissipation determines the direction of vortex
slippage, and gives $\vbf_v$ a component parallel to $\hat{r}$ for
$\omega>0$, and anti-parallel to $\hat{r}$ for $\omega<0$. For vortex
slippage at low dissipation angle, the mutual friction coefficients
satisfy
\begin{equation}
{\cal B}<< 1-{\cal B}^\prime <<1, 
\end{equation}
whereas many treatments of vortex pinning assume 
(\eg, \citealt{swc99,link06,gaj08a,vl08,ga09})
\begin{equation}
1-{\cal B}^\prime <<{\cal B} <<1. 
\end{equation}
This condition implicitly assumes very high drag, and is overly
restrictive.

\section{Mechanics and Energetics of Vortex Unpinning}

\label{AE}

Consider a vortex in stable equilibrium, pinned to a linear array of
equally-spaced pinning sites along the $z$ axis. In the presence of a
Magnus force, the vortex unpins by first assuming the unstable,
minimum-energy 
saddle-point configuration depicted in Figure \ref{saddlepoint}. In the
saddle-point configuration, a segment of vortex is free of $N_p$
pinning bonds. The activation energy $A$ is equal to the energy difference
between this state and the fully pinned state, and is determined by
the competition between the Magnus force, the pinning force, and the
self-energy (or tension) of the vortex.

Let the shape of the vortex be given by the
vector $\ubf(z)$. The energy of a vortex segment of length $L$ is
\citep{le91}
\begin{equation}
E=\int_L dz\, \left[
\frac{T_v}{2}\left|\frac{d\ubf(z)}{dz}\right|^2+
\rho_p V(u)
-\rho_s (\kappabf\times\vbf_s)\cdot\ubf(z)
\right].
\label{Ev}
\end{equation}
The first term is the energy cost of bending a vortex with a
self-energy per unit length $T_v$. The second term is the pinning
interaction, where $\rho_p=l_p^{-1}$ is the number of pinning sites per
unit length and $l_p$ is the average pinning spacing. The third term
is the potential energy associated with the constant Magnus force per
unit length.

\begin{figure*}[!ht]
\centering
\includegraphics[width=.6\linewidth]{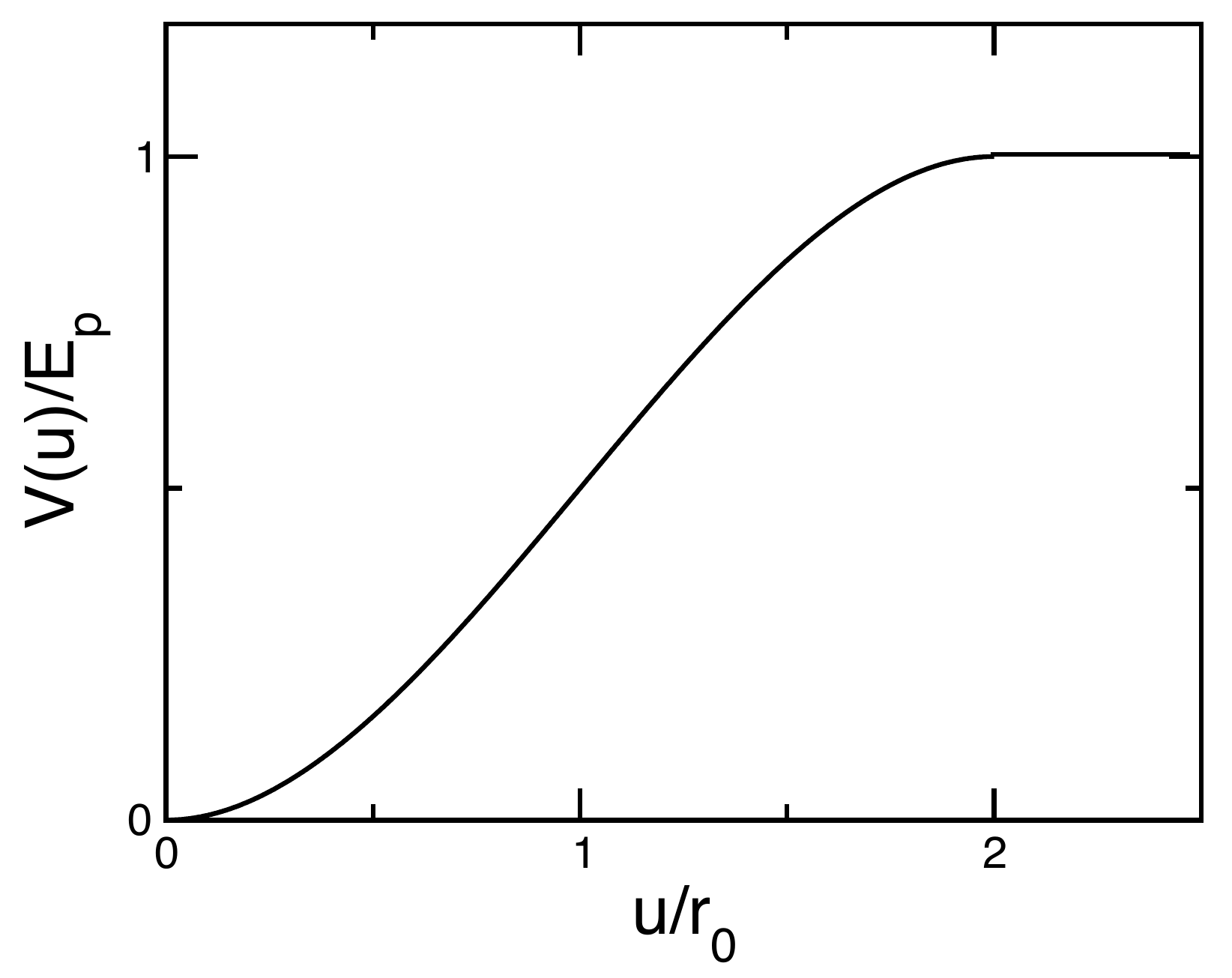} 
\caption{Potential of Equation [\ref{potential}], 
used by \citet{le91} to obtain the activation
energy of Equation [\ref{A}]. The perpendicular distance of the vortex
segment from a pinning site is $u$. The maximum force occurs for a
displacement $r_0$.}
\label{potentialplot}
\end{figure*}

The dependence of the interaction energy on the separation
between a vortex segment and a pinning site is unknown, but will
increase from zero to a maximum value $E_p$ over a length scale $r_0$. A
physically sensible parameterization for the potential is the
continuous piecewise function 
\begin{equation}
V(u)=
E_p\times 
\left\{\begin{array}{ll}
\frac{3}{4}\left(\frac{u}{r_0}\right)^2-\frac{1}{4}\left(\frac{u}{r_0}\right)^3
& 0\le u\le 2r_0 \\
1 & u>2r_0, 
\end{array}
\right.
\label{potential}
\end{equation}
shown in Figure \ref{potentialplot}.
The corresponding force per pinning site has the following
parabolic form:
\begin{equation}
F_p(u)=-\frac{dV}{du}=\left\{\begin{array}{ll}
-F_{\rm max}\left[1-\left(\frac{u}{r_0}-1\right)^2\right] & 
0\le u\le 2r_0 \\
0 & u>2r_0,
\end{array}
\right.
\end{equation}
where the maximum force is $F_{\rm max}=3E_p/4r_0$. The assumed
potential corresponds to the second-order series expansion of the
interaction force with a maximum of $F_{\rm max}$ at $r_0$, and so is
of general physical relevance. For pinning of vortices to nuclei in
the inner crust, the interaction range $r_0$ is $\sim \xi_n$, and the
pinning spacing is $l_p>a$.  For the pinning of vortices to flux tubes in
the outer core, the interaction range $r_0$ is $\sim \Lambda_*$, and
the pinning spacing is $l_p\sim l_\Phi$.

The large vortex self energy plays an
important role in determining the mechanics and energetics of
the unpinning process, quantified by the dimensionless tension parameter
\footnote{In the notation of \citet{le91}, $\tau={\cal T}$, the 
pinning energy is $U_0=E_p$, and  the pinning spacing is $l=l_p$.}
\begin{equation}
{\cal T}\equiv \frac{T_v r_0^2}{E_pl_p}. 
\end{equation}
If vortex tension were negligible, so that ${\cal T}$ is effectively
zero, the flexible vortex could unpin one pinning bond at a time to
assume a new, stable configuration. Because $T_v$ is non-zero, though,
the vortex must overcome $N_p>1$ pinning bonds in an unpinning event; 
$N_p$ is determined by ${\cal T}$. 
In the inner crust,
\begin{equation}
{\cal T}\simeq\frac{T_v \xi_n^2}{E_p l_p}\simeq 
0.31\,\rho_{n,14}\,E_p(\mbox{MeV})^{-1/2}\,\left(\frac{f_c}{0.1}\right)^{1/2}
\left(\frac{\xi_n}{10\mbox{ fm}}\right)^2 
\left(\frac{a}{50\mbox{ fm}}\right)^{-3/2}
\qquad
\mbox{(inner crust)},
\label{Tcal1}
\end{equation}
while in the outer core, 
\begin{equation}
{\cal T}\simeq \frac{T_v\Lambda_*^2}{E_pl_\Phi}
\simeq
0.12\,\left(\frac{\rho_{n,14}}{4}\right)
\left(\frac{E_p}{100\mbox{ MeV}}\right)^{-1} 
\left(\frac{\Lambda_*}{50\mbox{ fm}}\right)^2
B_{12}^{1/2}
\qquad
\mbox{(outer core)}.
\label{Tcal}
\end{equation}

Minimization of Equation [\ref{Ev}] gives the stable and saddle-point 
vortex configurations
and the associated energies; see \citet{le91} for details. 
For a linear array of pinning sites with the potential of
Equation [\ref{potential}], 
the number of broken pinning bonds in the saddle-point configuration
is 
\begin{equation}
N_p\simeq 3.2{\,\cal T}^{1/2}(1-\omega/\omega_{\rm crit})^{-1/4}, 
\end{equation}
and the activation energy is
\begin{equation}
A\simeq 1.6 N_p E_p (1-\omega/\omega_{\rm crit})^{3/2}
\simeq
5.1 E_p {\cal T}^{1/2} (1-\omega/\omega_{\rm crit})^{5/4}
\label{A}
\end{equation}
These expressions are valid for $3/4\le\omega/\omega_{\rm crit}< 1$
and $N_p\gap 3$. The latter condition gives 
$1-\omega/\omega_{\rm crit}<{\cal T}^2$; 
for typical parameters of the inner crust and the outer core, $\omega$
will always be very close to $\omega_{\rm crit}$, so that
this condition is satisfied. Hence, in both the inner crust and
the outer core $N_p>>1$, and the vortex is effectively stiff for all pinning
energies of interest. Vortex stiffness becomes more important as $E_p$
is reduced. The important effects of vortex stiffness in determining
the activation energy were not considered by AAPS or in subsequent
development of vortex creep theory.

To see the basic scaling of $N_p$ and $A$ with ${\cal T}^{1/2}$, note
that the energy of a vortex segment bent by amplitude $u$ is, from 
Equation [\ref{Ev}], 
\begin{equation}
E\sim \frac{1}{2}T_v\frac{u^2}{L_p}+\frac{L_p}{l_p}E_p, 
\end{equation}
for zero Magnus force. 
Extremizing the energy in $L_p$ gives
\begin{equation}
E\sim \left(\frac{2T_vu^2}{E_pl_p}\right)^{1/2}E_p.
\label{Eest}
\end{equation}
If the vortex is to unpin, it must be excited to an amplitude
comparable to the range of the pinning potential,
$u\sim r_0$. Hence, 
\begin{equation}
A\sim (2{\cal T})^{1/2}E_p\sim N_pE_p, 
\end{equation}
where $N_p=L_p/l_p$ is the characteristic number of bonds set by
vortex tension that must
be broken for the vortex to unpin. 

\section{Relaxation Dynamics with Vortex Slippage}

\label{relaxation}

Suppose a glitch occurs at $t=0$, through some instability that 
need not be specified for this perturbative analysis, and let us
consider the dynamics in the $C+S+L$ model of
Equation [\ref{jcons}]. For axisymmetric rotation, the dynamics is given by
\begin{equation}
I_c\dot\Omega_c+\int
dI_s\,\dot\Omega_s=-I\vert\dot\Omega_0\vert+J_l\,\delta(t), 
\label{Jcons7}
\end{equation}
\begin{equation}
\dot\Omega_s(r,t)=\left(\Omega_s(r,t)+\frac{r}{2}\frac{\partial\Omega_s(r,t)}
{\partial r}\right)\omega(r,t)\sin2\theta_d\, {\rm e}^{-\beta
A(\omega(r,t))},
\label{Omsdot_full}
\end{equation}
where $r$ is the cylindrical radius. The response of the crust is
determined by integrated dynamics of the superfluid on cylindrical
shells of radius $r$. 

The system can be reduced to a single integral
equation with some approximations. 
The second term in parenthesis in Equation [\ref{Omsdot_full}] is 
\begin{equation}
\frac{r}{2}\frac{\partial\Omega_s(r,t)}{\partial r}\simeq
\frac{r}{2}\frac{\partial\omega_{\rm crit}(r))}{\partial r}, 
\end{equation}
where, from Equations [\ref{omc_crust}] and [\ref{omc_core}], $\omega_{\rm
crit}\propto r^{-1}$. Hence, 
\begin{equation}
\frac{r}{2}\frac{\partial\Omega_s(r,t))}{\partial r} \simeq
-\frac{1}{2}\omega_{\rm crit}(r).
\end{equation}
The critical lag $\omega_{\rm crit}$ is always much smaller than
$\Omega_s$, so the second term in Equation [\ref{Omsdot_full}]
can be neglected, giving
\begin{equation}
\dot\Omega_s(r,t)=-\Omega_s(r,t)\,\omega(r,t)\,\sin 2\theta_d\, {\rm e}^{-\beta
A(\omega(r,t))}.
\end{equation}
Expanding $A(\omega)$ about $\omega_0$ in the exponential gives 
\begin{equation}
\dot{\Omega}_s(r,t)\simeq\dot{\Omega}_0\,{\rm e}^{b(\omega(r,t)-\omega_0(r))}
\qquad
b(r)\equiv \left.-\beta\,\frac{\partial
A}{\partial\omega}\right|_{\omega_0(r)}.
\label{eom}
\end{equation}
Combining this equation with Equation [\ref{Jcons7}] gives the equation of motion
for the lag at each polar radius $r$, for $t>0$
\begin{equation}
\dot\omega(r,t)=
\vert\dot\Omega_0\vert\left(1-{\rm
e}^{b(r)(\omega(r,t)-\omega_0(r))}\right)
+\frac{1}{I_c}\int dI_s\,
\vert\dot\Omega_0\vert\left(1-{\rm
e}^{b(r)(\omega(r,t)-\omega_0(r))}
\right). 
\label{lageom}
\end{equation}
Let us assume that the regions with pinning constitute a small
fraction of the star's moment of inertia, so that $I_s<<I_c$. In this
limit the spatial integral in the above equation can be
neglected. With these approximations, the evolution of the lag is determined
entirely by its value locally, and the response of the crust can be
calculated by integrating $\omega(r,t)$ over cylindrical shells. These
simplifications were introduced by \citet{alpar_etal84a}. 

Ignoring the spatial integral in Equation [\ref{lageom}], the solution is 
\citep{alpar_etal84a,leb93}
\begin{equation}
\dot{\omega}(r,t)=\vert\dot{\Omega}_0\vert
\left[1-\frac{1}{1+({\rm e}^{t_d(r)/t_r(r)}-1){\rm e}^{-t/t_r(r)}}\right]. 
\label{solution}
\end{equation}
where 
\begin{equation}
t_d(r)\equiv\frac{\omega_0(r)-\omega(r,0+)}
{\vert\dot{\Omega}_0\vert} 
\qquad\qquad
t_r(r)\equiv \frac{1}{b(r)\vert{\dot{\Omega}_0\vert}}
\label{tt}
\end{equation}
Here $t_d(r)$ is the {\em decoupling time} of the shell at radius $r$ 
and $t_r$ is the intrinsic {\em recoupling time} of the
shell. In the limit $t_d>>t_r$, a perturbation in
$\omega$ leaves components $C$ and $S$ decoupled for a time $\sim
t_d$, before the perturbation is damped over a recoupling time $t_r$. 
In the opposite limit $t_d<<t_r$, a perturbation damps
over a time $t_r$. This behavior is a consequence of the non-linear
dependence of the slippage velocity on the initial value of
$\omega$. Equation [\ref{solution}] was also obtained by
AAPS for the special case of the activation energy
having linear dependence on $\omega$; the form given here is more
general with the quantity $b$ given in Equation [\ref{eom}].

As a simplification, let us specify the pre-glitch state by assuming
that all transient rotational dynamics has damped before the glitch,
so that the system is in in {\em spin-down equilibrium} at $t=0$, such
that $\dot{\omega}=0$, $\dot{\Omega}_s=\dot{\Omega}_c=\dot{\Omega}_0$
for lag $\omega_0$. We will find that the post-glitch recovery time of
the inner crust greatly exceeds the average interval between glitches,
so rotational equilibrium in the inner crust is not reached between
glitches (more on this point in Section \ref{timescales}). By contrast, we
will find that the core has ample time to recover rotational
equilibrium between glitches.  

The equilibrium lag $\omega_0$ is given
by the solution to
\begin{equation}
\dot{\Omega}_0=-\Omega_s(r)\,\omega_0(r)
\sin2\theta_d\, {\rm e}^{-\beta A(\omega_0(r))}.
\label{equilibrium}
\end{equation}
The residual spin rate of the crust is given by Equation [\ref{resid}]:
\begin{equation}
\delta\Omega_c(t)=-\frac{1}{I}\int dI_s\, 
\left[
\omega(r,t)-\omega_0(r)\right]+\frac{J_l}{I},
\end{equation}
To determine the form and timescale of the response, I assume for
simplicity that the glitch is driven by component $L$, so that 
\begin{equation}
\omega(r,0+)=\omega_0(r)+\Delta\Omega_c.
\end{equation}
The decoupling time is independent of both $r$ and the details of pinning:
\begin{equation}
t_d=\frac{\Delta\Omega_c}{\vert\dot\Omega_0\vert}=2t_{sd}\frac{\Delta\Omega_c}{\Omega_c},
\label{td}
\end{equation}
where $t_{sd}\equiv\Omega_c/2\vert\dot\Omega_c\vert$ is the spin-down
age. 
The spin rate residual $\delta\Omega_c$, in units of the initial spin
jump $\Delta\Omega_c$, is
\begin{equation}
\frac{\delta\Omega_c(t)}{\Delta\Omega_c}=
1-\frac{1}{I}\int dI_s\,
\left[\frac{t}{t_d}-\int_{0+}^t\frac{dt^\prime}{t_d}\,
\frac{1}{1+({\rm e}^{t_d/t_r(r)}-1){\rm e}^{-t^\prime/t_r(r)}}\right]. 
\label{fullsolution}
\end{equation}
At late times, the time integral approaches unity, giving
\begin{equation}
\lim_{t\rightarrow\infty}\frac{\delta\Omega_c(t)}{\Delta\Omega_c}=
\frac{I_c}{I}=1-Q;
\end{equation}
the glitch recovers by a fraction $Q=I_s/I$. 

Now consider the pinning region to be in a  spherical shell of constant
density $\rho$ and thickness $\Delta$, inner radius $R$, and outer radius
$R+\Delta$. The height of the shell in each hemisphere is $h(r)$. The
response of the crust is given by 
\begin{equation}
\frac{\delta\Omega_c(t)}{\Delta\Omega_c}=
1-\frac{1}{I} \int_0^{R+\Delta}dr\, 2\pi\rho h(r) r^3\,
\left[
\frac{t}{t_d}-
\int_{0+}^t
\frac{dt^\prime}{t_d}\,
\frac{1}{1+({\rm e}^{t_d/t_r(r)}-1){\rm e}^{-t^\prime/t_r(r)}}
\right], 
\label{fullsolution1}
\end{equation}
where 
\begin{eqnarray}
h(r)= \left\{
\begin{array}{ll}
\sqrt{(R+\Delta)^2-r^2}-\sqrt{R^2-r^2} & r<R \\
\sqrt{(R+\Delta)^2-r^2} & r\ge R
\end{array}
\right.
\end{eqnarray}

\begin{figure*}[t]
\centering
\includegraphics[width=.6\linewidth]{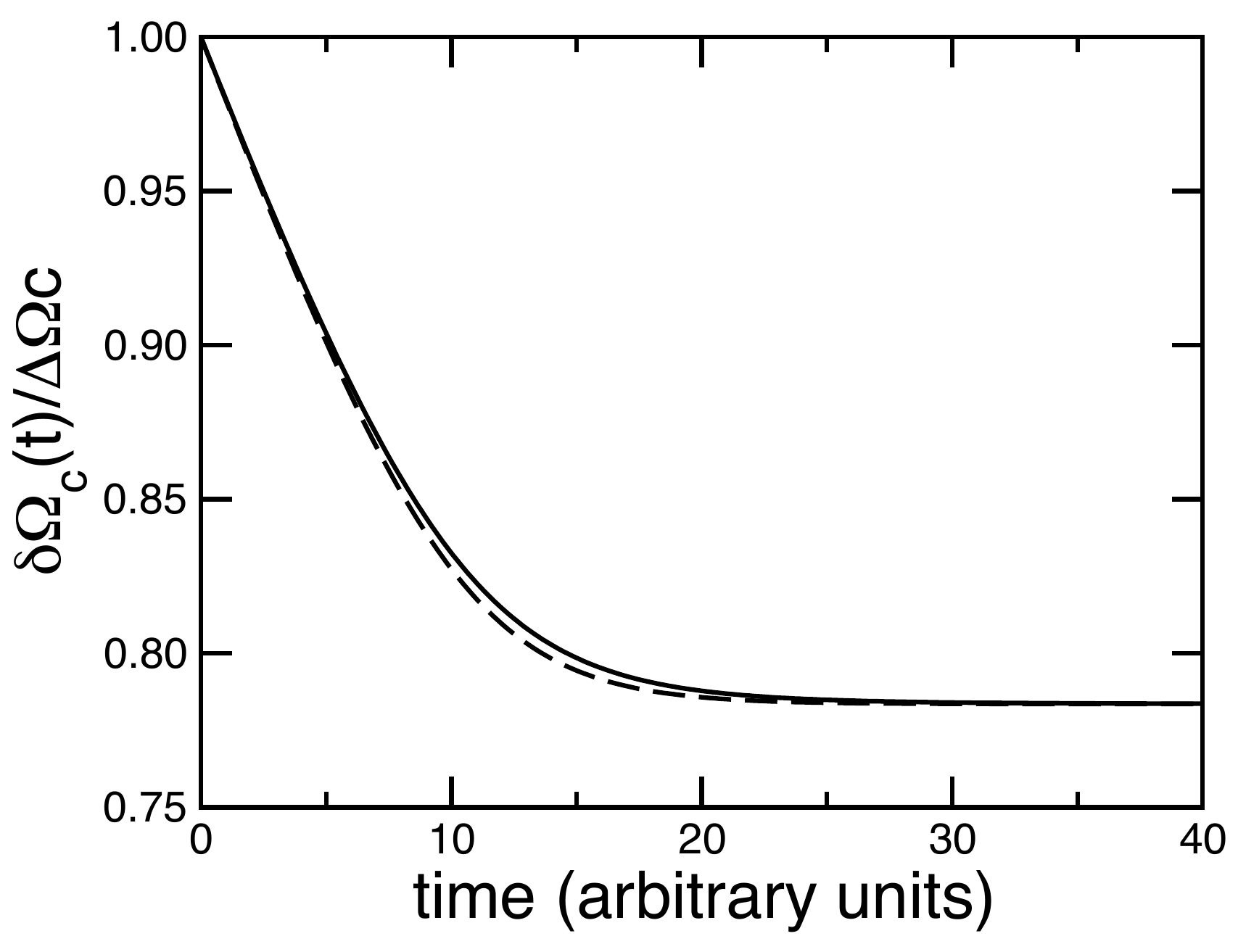} 
\caption{Solution given by Equation [\ref{fullsolution1}] (solid curve)
compared to
the simpler solution of Equation [\ref{fullsolution2}] (dashed curve), for
$\Delta/R=0.05$, $I_s/I=0.22$, $t_d=10$, and $t(R)=3$.}
\label{shellmodel}
\end{figure*}

We will see that $t_r$ scales as $\omega_{\rm crit}\propto r^{-1}$.
Cylindrical shells with smaller radii thus have the longest recoupling
times, but they also have smaller moments of inertia $\propto
r^4\Delta$. The response will be largely determined by the relaxation
time in the region $R<r<R+\Delta$. Let the characteristic recoupling 
time there be $t_r(R)$. Equation [\ref{fullsolution1}] then becomes
\[
\frac{\delta\Omega_c(t)}{\Delta\Omega_c}\simeq 
1-\frac{I_s}{I} 
\left[
\frac{t}{t_d}-
\int_{0+}^t \frac{dt^\prime}{t_d}\,
\frac{1}{1+({\rm e}^{t_d/t_r(R)}-1){\rm e}^{-t^\prime/t_r(R)}}
\right]=
\]
\begin{equation}
1-\frac{{I_s}}{t_dI}\left[t
+t_r(R)\ln\left(\frac{1+c(R)}{{\rm e}^{t/t_r(R)}+c(R)}\right)
\right] 
\qquad
c(R)\equiv {\rm e}^{t_d/t_r(R)}-1.
\label{fullsolution2}
\end{equation}
Figure \ref{shellmodel} shows that Equation [\ref{fullsolution2}] is a good 
approximation to Equation [\ref{fullsolution1}]. Henceforth I write
$t_r(R)=t_r$. 

Now suppose we add concentric spherical shells, each with superfluid moment of
inertia $I_{s,i}$ and characteristic relaxation time $t_r(R_i)\equiv t_{r,i}$. 
Because of the additive nature of the contributions to the post-glitch
recovery within the approximations made so far, the total response of
the star is 
\begin{equation}
\frac{\delta\Omega_c(t)}{\Delta\Omega_c}\simeq 
1-\frac{1}{t_dI}\sum_i I_{s,i}
\left[t
+t_{r,i}\ln\left(\frac{1+c_i}{{\rm e}^{t/t_{r,i}}+c_i}\right)
\right]
\qquad
c_i\equiv {\rm e}^{t_d/t_{r,i}}-1,
\label{fullsolution3}
\end{equation}
and 
\begin{equation}
Q=\frac{1}{I}\sum_i I_{s,i}.
\end{equation}
For simplicity I continue to consider the response of a star with a single
pinning zone, since the results are easily generalized for multiple
pinning zones. 

In general, the response of the crust has non-linear dependence on the
size of the glitch through the appearance of $t_d$, and is referred to
as the ``non-linear creep regime'' by \citet{alpar_etal89}. The
integral of Equation [\ref{fullsolution2}] can be evaluated analytically
in two limits. If the glitch is small enough that $t_d<<t_r$,
expanding the integrand to first order in $t_d/t_r$ gives
\begin{equation}
\frac{\delta\Omega_c(t)}{\Delta\Omega_c}=1-Q(1-{\rm e}^{-t/t_r})
\qquad
t_d<<t_r,
\label{exponential}
\end{equation}
as for the case for linear frictional coupling (Equation \ref{friction}),
but now with a restriction on the glitch magnitude; the response to
the glitch consists of {\em exponential relaxation} over a
timescale $t_r$. 
The residual in
the spin-down rate, $\delta\dot{\Omega}_c(t)$, is
\begin{equation}
\frac{\delta\dot{\Omega}_c(t)}{\vert\dot{\Omega}_0\vert}=
-Q\frac{\Delta\Omega_c}{\vert\dot{\Omega}_0\vert t_r}{\rm
e}^{-t/t_r}.
\end{equation}
For a single $S$ component, the glitch-induced discontinuity in the
spin-down rate is 
\begin{equation}
\frac{\delta\dot{\Omega}_c(0)}{\vert\dot{\Omega}_0\vert}=
-\frac{I_s}{I}\frac{\Delta\Omega_c}{\vert\dot{\Omega}_0\vert}
\frac{t_{sd}}{t_r}.
\end{equation}

\begin{figure*}[!ht]
\centering
\includegraphics[width=.6\linewidth]{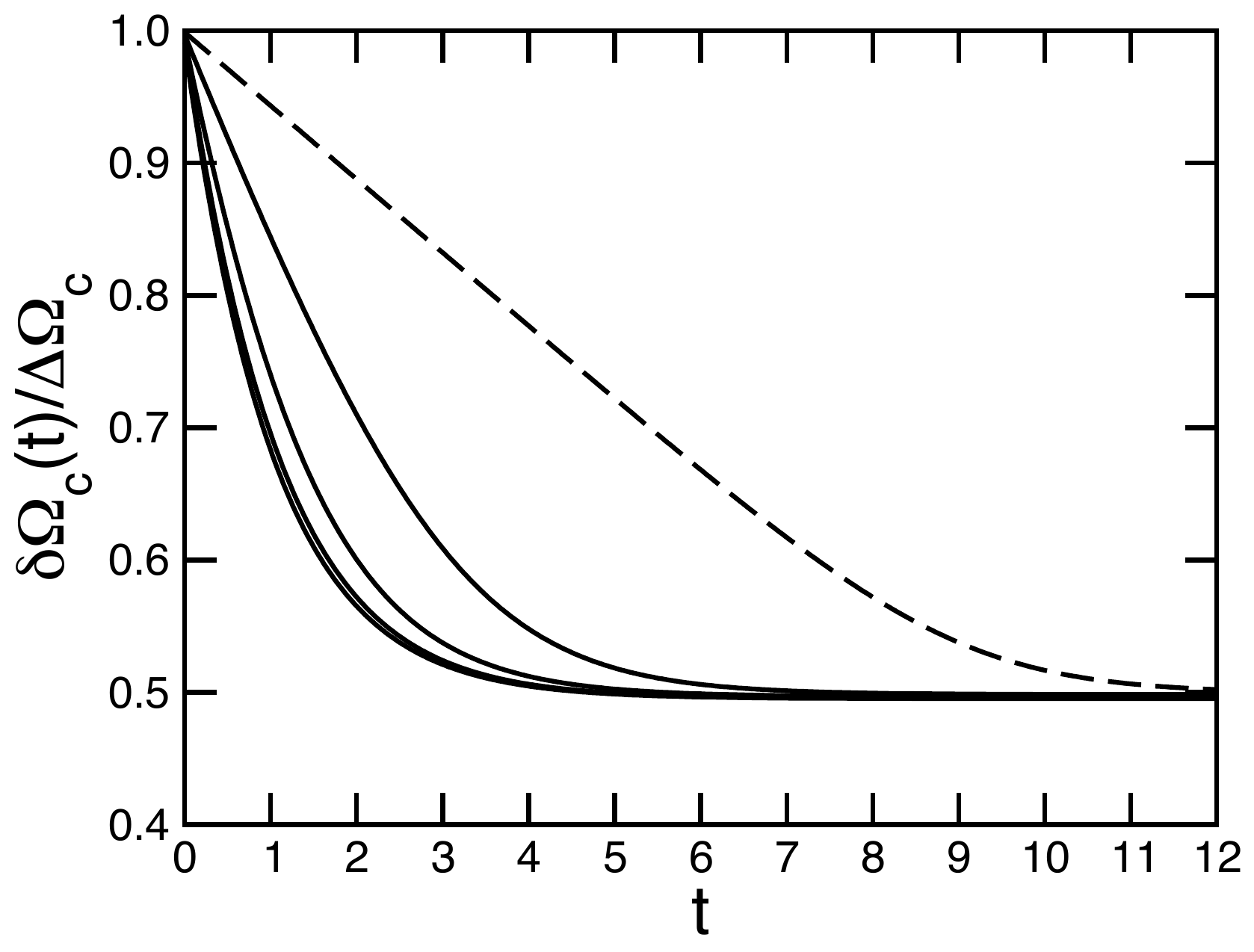} 
\caption{Recovery of the crust's spin rate to a glitch for $Q=0.5$,
$t_r=1$, and 
(left to right) $t_d=0.1,0.3,1,3,9$. For $t_d>>t_r$, the response
consists of linear decay over a time $t_d$, followed by quick
recovery to $1-Q$ over a timescale $t_r$. For $t_r>>t_d$, the
response is exponential decay over a timescale $t_r$. The total
relaxation time is $\tau=t_d+t_r$. The dashed curve shows the
distinctive signature of recovery of a large glitch through thermal
activation of pinned vortices. Significant deviations from exponential
response occur for $t_d\gap 3t_r$, corresponding to $\tau\gap t_d$. }
\label{response_resid}
\end{figure*}

\begin{figure*}[!ht]
\centering
\includegraphics[width=.6\linewidth]{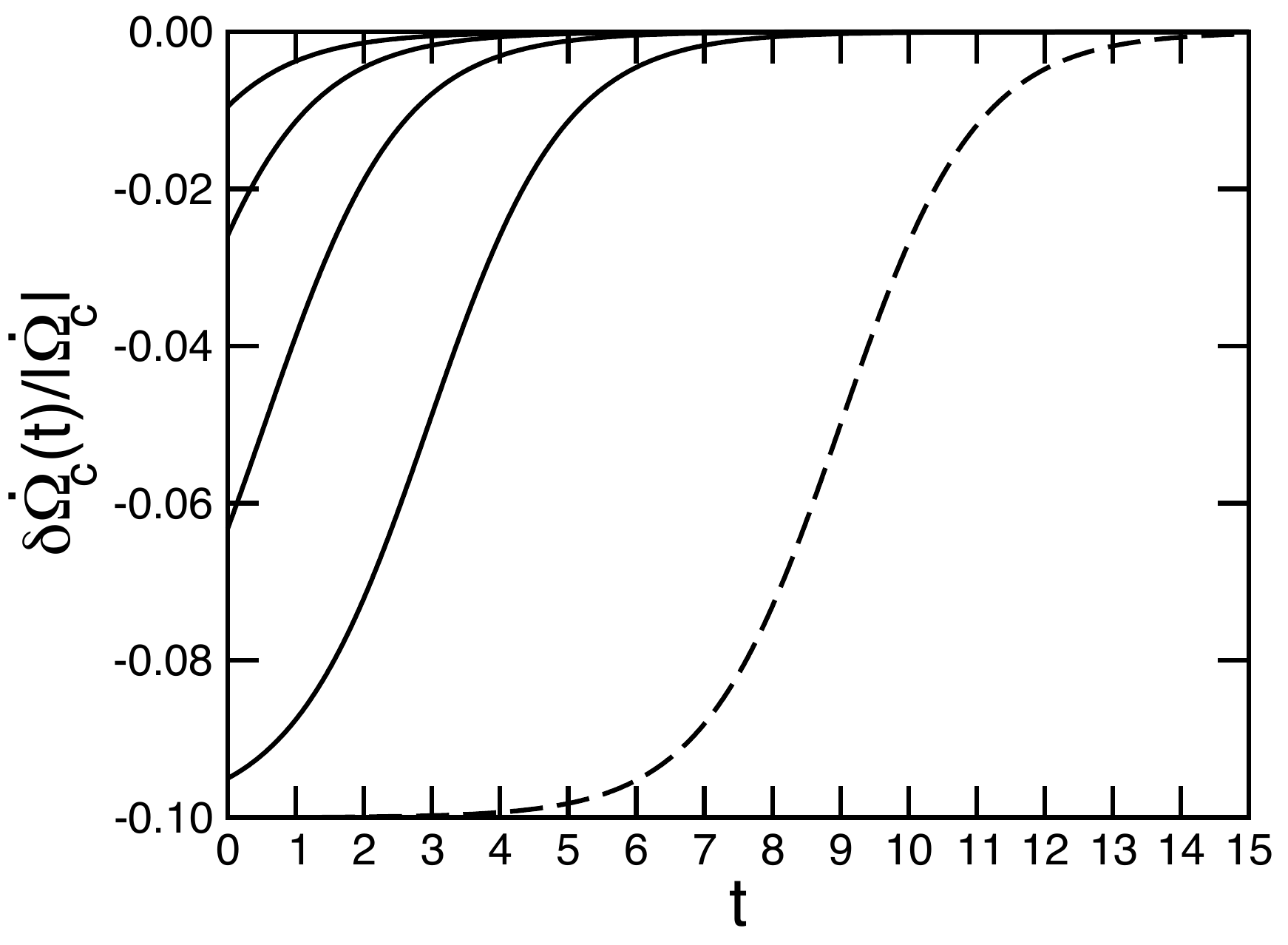} 
\caption{Recovery of $\delta\dot{\Omega}_c$ for $I_s/I_c=0.1$, 
$t_r=1$, and
(left to right) $t_d=0.1,0.3,1,3,9$. For $t_d>>t_r$, $S$ is decoupled
from $C$ for a time $t_d$, giving a fractional decrease of $-I_s/I_c$
in $\vert\delta\dot{\Omega}_c\vert$, before rotational equilibrium is
restored over a time $\simeq t_d$. The dashed curve shows the
distinctive signature of recovery of a large glitch through thermal
activation of pinned vortices.
For $t_r>>t_d$, the recovery consists of
exponential decay, with a fractional change in
$\delta\dot{\Omega}_c$ at $t=0+$ that is not related to $I_s/I_c$. 
For fixed $Q$, the glitch magnitude and $\delta\dot\Omega_c(0+)$
both decrease with $t_d$ (Equation \ref{td}).}
\label{response_dresid}
\end{figure*}

In the opposite limit of a large glitch, so that $t_d>>t_r$, the
integrand of Equation [\ref{fullsolution2}] becomes a Fermi function 
\begin{equation}
\frac{1}{1+({\rm e}^{t_d/t_r}-1){\rm e}^{-t^\prime/t_r}}\simeq
\frac{1}{1+{\rm e}^{-(t^\prime-t_d)/t_r}}, 
\end{equation}
and the response of the crust is 
\begin{equation}
\frac{\delta\Omega_c(t)}{\Delta\Omega_c}=
1-Q\left[\frac{t}{t_d}
-\frac{t_r}{t_d}\ln\left(1+{\rm e}^{(t-t_d)/t_r}\right)\right]
\qquad
t_d>>t_r, 
\label{nonexponential}
\end{equation}
The star recovers through {\em delayed response}, as shown by the
dashed curve in 
Figure \ref{response_resid}. 
The residual in the spin-down rate is
\begin{equation}
\frac{\delta\dot{\Omega}_c(t)}{\vert\dot{\Omega}_0\vert}=
-Q\frac{\Delta\Omega_c}
{\vert\dot{\Omega}_0\vert}\frac{1}{1+{\rm e}^{(t-t_d)/t_r}}=
-\frac{I_s}{I_c}\frac{1}{1+{\rm e}^{(t-t_d)/t_r}} \qquad t_d>>t_r.
\label{eq87}
\end{equation}
Right after the glitch, the magnitude of the spin rate has increased
by $\vert\delta\dot\Omega_c\vert/\vert\dot\Omega_c\vert=I_s/I_c$. This
constant spin rate excess persists for a time $t_d$, followed by
relatively quick recovery over a timescale $t_r$, as shown by the
dashed curve in Figure \ref{response_dresid}. Examples of the general
behavior of $\delta\Omega_c(t)$ and $\delta\dot{\Omega}_c(t)$ are
shown in Figure \ref{response_resid} and \ref{response_dresid}. The
characteristic relaxation time of the system is $\tau=t_d+t_r$.
Significant deviations from exponential
response occur for $t_d\gap 3t_r$, corresponding to $\tau\gap t_d$.

The recoupling time $t_r$ depends 
on the value of the equilibrium lag $\omega_0$. 
The equilibrium lag is given by the 
solution to Equations [\ref{equilibrium}] and [\ref{A}]:
\begin{equation}
1-\frac{\omega_0}{\omega_{\rm crit}}=
\left[
\frac{kT}{5.1{\cal T}^{1/2}E_p}\ln\left(2t_{sd}\,\omega_0\sin
2\theta_d\,\right)\right]^{4/5}, 
\label{eqlag}
\end{equation}
where $\Omega_s\simeq\Omega_c$ was used. For typical parameters, the
right-hand side of Equation [\ref{eqlag}] is much smaller than unity, so
that $\omega_0\simeq\omega_{\rm crit}$; $\omega_0$ can be replaced by
$\omega_{\rm crit}$ to a very good approximation. 
Equations [\ref{tt}], [\ref{eom}], and [\ref{A}] give the recoupling time
\begin{equation}
t_r=0.22\frac{I_c}{I}\omega_{\rm crit}\vert\dot{\Omega}_0\vert^{-1}
({\cal T}^{1/2}\beta E_p)^{-4/5}
\left[\ln(2t_{sd}\omega_{\rm crit}\,\sin 2\theta_d)\right]^{-1/5}.
\label{trfinal}
\end{equation}
Note the extremely weak dependence on $t_{sd}$ and $\theta_d$, and
linear dependence on $\vert\dot\Omega_0\vert^{-1}$. Unlike $t_d$, the
recoupling time $t_r$ does not depend on the conditions after the glitch;
$t_r$ is intrinsic to the system (or part of the system), and
does not depend the simple $C$+$S$+$L$ model used to calculate $t_d$. 

The chief mathematical results of the vortex slippage theory developed
in this paper are given by Equations [\ref{td}], [\ref{fullsolution3}], and
[\ref{trfinal}]. I now proceed with numerical estimates. 

\section{Estimates of Post-glitch Recovery Timescales}

\label{timescales}

The decoupling time in regions of star in which there is no vortex
motion at the time of the glitch is, from
Equation [\ref{td}], 
\begin{equation}
t_d=7\,\left(\frac{t_{sd}}{10^4\mbox{ yr}}\right)
\left(\frac{\Delta\nu/\nu}{10^{-6}}\right)
\mbox{ days}
\label{tdest}
\end{equation}
where $\nu$ is the spin frequency of the pulsar and $\Delta\nu$ is the
change at the glitch.
For large glitches in older stars, $t_d$ can
be very long; $\Delta\nu/\nu=3\times 10^{-6}$ gives $t_d=210$ d for
$t_{sd}=10^5$ yr. In the region or regions of the star in which vortex
motion drives the glitch, $t_d$ can be much longer; see below. 

The recoupling time $t_r$ is determined by the pinning parameters. For
vortex slippage in the inner crust, $t_r$ follows from
Equations [\ref{trfinal}], [\ref{lp}], [\ref{omc_crust}], and [\ref{Tcal1}]. Taking
$\sin 2\theta_d=10^{-2}$ for dissipation through Kelvin phonon production
(though the exact value is unimportant), $t_r$ in terms of fiducial values is 
\begin{equation}
t_r\simeq 30\, 
\left(\frac{\vert\dot{\nu}\vert}{10^{-11}\mbox{ s$^{-2}$}}\right)^{-1}
\left(\frac{f_c}{0.1}\right)^{-1.7}\left(\frac{\rho_{n,14}}{0.5}\right)^{-1.7}
E_p(\mbox{MeV})^{0.9}\times
\left(\frac{kT}{10\mbox{ keV}}\right)^{4/5}
\label{trcrust}
\end{equation}
\[
\hspace*{3.cm}
\left(\frac{kT}{10\mbox{ keV}}\right)^{4/5}
\left(\frac{a}{50\mbox{ fm}}\right)^{-0.9}
\left(\frac{\xi_n}{10\mbox{ fm}}\right)^{-9/5}
\left(\frac{R}{10\mbox{ km}}\right)^{-1}
\mbox{ yr}.
\]
The population of glitching pulsars shows a range in $\dot{\nu}$ from
$\sim -10^{-15}$ s$^{-2}$ to $\sim -10^{-9}$ s$^{-2}$, giving values
of $t_r$ from $\sim 10^2$ d to decades. Nuclear entrainment has the
important effect of making $t_r$ very long by reducing the density of
free neutrons. The Magnus force (for given lag) that
drives the system's recovery to rotational equilibrium is
correspondingly reduced, so the recovery time is increased. 

Recoupling of the core through vortex slippage occurs much more
quickly than in the inner crust. Using Equations [\ref{lphi}], 
[\ref{omc_core}], and [\ref{Tcal}] in
Equation [\ref{trfinal}], and taking fiducial values for the outer
core, gives:
\[
t_r\simeq 3\,
\left(\frac{\vert\dot\nu\vert}{10^{-11} \mbox{ s$^{-2}$}}\right)^{-1}
\left(\frac{\rho_{n,14}}{4}\right)^{-7/5} 
\left(\frac{kT}{10\mbox{ keV}}\right)^{4/5} 
\left(\frac{E_p}{100\mbox{ MeV}}\right)^{3/5}\times
\hspace*{3.cm}
\]
\begin{equation}
\left(\frac{\Lambda_*}{50\mbox{ fm}}\right)^{-9/5}
\left(\frac{R}{10\mbox{ km}}\right)^{-1}
B_{12}^{3/10}
\mbox{ days}.
\label{trcore}
\end{equation}
While the dipole components of the field typically have $B_{12}=1$, it
is the internal field that determines the recoupling time. With
relatively weak sensitivity to $E_p$ and $B$, this estimate is more
robust than for the inner crust. Higher multipoles could plausibly
make $B$ about ten times larger than the surface field
\citep{braithwaite09}.

For large glitches, $t_r$ is typically less than $t_d$, and so the
response will be the delayed response illustrated by the dashed curves of
Figures \ref{response_resid} and \ref{response_dresid}; see the
discussion following Equation [\ref{eq87}].  If a moment of
inertia $\Delta I_s$ is decoupled by the glitch, post-glitch response
of the core consists of a fixed increase in the magnitude of the
spin-down rate by a factor $\Delta I_s/I_c$, followed by a relatively
quick recovery over a timescale $t_r$, after time $t_d$.

The large vortex self-energy acts to reduce the recoupling time
$t_r$; see Equation [\ref{trfinal}]. Increasing the
tension by a factor of 10 over that used in Equation [\ref{tension}]
decreases $t_r$ by a factor of $\simeq 5$ in the
inner crust and $\simeq 3$ in the outer core. Vortex tension
generally increases vortex mobility, giving shorter recoupling
times. The decoupling time $t_d$ is unaffected. 

The estimate of $t_d$ in Equation [\ref{tdest}] applies to all parts of the
star in which pinned vortices do not move during the glitch. 
If some portion $S^\prime$ of the superfluid interior with a moment of
inertia $I_s^\prime$ drives the glitch, and $I_s^\prime$ is much less
than $I_c$, the decoupling time of this region could be very
long. By angular momentum conservation
\begin{equation}
I_c\Delta\Omega_c + I_s^\prime \Delta\Omega_s=0, 
\end{equation}
giving a change in lag in this region given by 
\begin{equation}
I\Delta\Omega_c+I_s^\prime\Delta\omega=0,
\end{equation}
and a decoupling time of this region of 
\begin{equation}
t_d^\prime=\frac{\vert\Delta\omega\vert}{\vert\dot\Omega_c\vert}=
\frac{I}{I_s^\prime}\frac{\Delta\Omega_c}{\vert\dot\Omega_0\vert}
=2\, \left(\frac{I/I_s^\prime}{10^{-2}}\right)
\left(\frac{t_{sd}}{10^4\mbox{ yr}}\right)
\left(\frac{\Delta\nu/\nu}{10^{-6}}\right)
\mbox{ yr}
\end{equation}
Such a region could serve as the $L$ component in the simple model of
this paper as follows. After the glitch, $S^\prime$ remains decoupled
from $C$ for a time $t_d^\prime$, and stores angular momentum as the
star spins down. After a time $t_d^\prime$, $S^\prime$ recovers to
spin-down equilibrium over the recoupling time $t_r^\prime$ of that
region. An instability drives vortex motion in $S^\prime$, creating a
glitch, and the process repeats itself. Parts of the inner crust with
typically very long coupling times, could serve as the $L$
component. 

For all other parts of the star in which there is pinning, $t_d$ is
the same, and is set by the size of the glitch according to
Equation [\ref{tdest}].

\section{Comparison with observed post-glitch response}

\label{data}

I now give a brief comparison for two pulsars for which post-glitch
response has been measured in detail: the Vela and Crab pulsars. Vela
has shown post-glitch response times that range between about 0.5 d and about
400 d. The Crab pulsar has shown post-glitch response times that range
from about two days to 100 d. Most glitches show multiple
timescales; see the ATNF Pulsar Glitch Catalogue
\citep{manchester_etal05} and references therein. 

\subsection{Inner-crust slippage}

Equation [\ref{trcrust}] predicts $t_r\simeq 10^4$ d for the Vela pulsar,
with a relatively small addition to the total relaxation of $t_d\simeq
20$ d for a typical glitch. For the Crab pulsar the predicted
recoupling time is $t_r\simeq 300$ d, with $t_d<<t_r$ because the
glitches are generally small. It appears that post-glitch response due
to vortex slippage in the inner crust is too slow to reconcile with
observations in these two pulsars. Note that $t_r$ decreases by a
factor of about 50 if entrainment is negligible for some reason. 

\subsection{Outer-core crust slippage}

For the Vela pulsar, the total relaxation time $t_d+t_r$ from
Equations [\ref{tdest}] and [\ref{trcore}] is not in significant
disagreement with observed relaxation times but the predictions for
the Crab pulsar are $\tau<0.1$ d, far shorter than observed; details
will be given in a subsequent publication. 

\bigskip

These preliminary comparisons suggest that most (if not all) aspects
of post-glitch response are not due to thermally-activated vortex
activation, but arise from a different process. The dominant process
in post-glitch response might be vortex drag in regions were there is
no pinning.

\section{Comparison with Earlier Work}

\label{comparison}

The theory developed here differs significantly from the 
vortex creep theory of AAPS applied to the inner crust
(see, also,
\citealt{alpar_etal84b,alpar_etal85,alpar_etal86,alpar_etal88,alpar_etal89,alpar_etal93,alpar_etal94,alpar_etal96})
and the extensions of the AAPS theory to vortex creep in the outer
core \citep{sa09}, and gives different predictions.
The differences are due to different treatments
of the vortex velocity and different forms for the
activation energy. 

AAPS and \citet{alpar_etal84b} adopt the following expression for the
radial component of the creep velocity: 
\begin{equation}
\langle\vbf_v\rangle\cdot\hat{r}=v_0{\rm e}^{-\beta E_p(1-\omega/\omega_{\rm crit})}
-v_0{\rm e}^{-E_p\beta (1+\omega/\omega_{\rm crit})}
=2v_0 {\rm e}^{-\beta E_p}\sinh\left(\beta
E_p\frac{\omega}{\omega_{\rm crit}}\right); 
\label{valpar}
\end{equation}
{\sl c.f.}, Equation [\ref{vslippage}].  The first term accounts for motion
away from the rotation axis, and the second term for motion toward the
rotation axis. The factor $v_0$ is a velocity that AAPS refer to as
``a typical velocity of microscopic motion of the vortex lines between
pinning centers''; a calculation of $v_0$ is not given, and $v_0$ is
fixed at a constant value of $10^7$ cm s$^{-1}$ for both the inner
crust and outer core. The activation energies for outward (+) and
inward (-) vortex motion are assumed to be
\begin{equation}
A_\pm=E_p(1\mp\omega/\omega_{\rm crit}). 
\label{Aalpar}
\end{equation}
This form assumes that vortices are moving in a {\em linear} bias
imposed by the average Magnus force; {\sl c.f.}, Equation [\ref{A}]. Note
that vortex tension does not appear in Equations [\ref{valpar}] and
[\ref{Aalpar}]; each pinned segment of a vortex is treated effectively
as a point particle, without coupling of adjacent segments through
tension. 

The chief conclusions of the vortex creep theory of AAPS and its
extensions are these:

\begin{enumerate}

\item For relatively large $E_p$, spin-down drives the system to an
equilibrium lag 
$\omega_0\lap\omega_{\rm crit}$. The radial component of the creep
velocity becomes, in the theory of AAPS:
\begin{equation}
\langle\vbf_v\rangle\cdot\hat{r}=v_0{\rm e}^{-\beta
E_p(1-\omega/\omega_{\rm crit})} \qquad \omega_0\lap\omega_{\rm crit}.
\label{valpar_nonlinear}
\end{equation}
In the nomenclature of
\citet{alpar_etal89}, this limit is the regime of ``non-linear
creep''. 
In this regime, AAPS obtain a decoupling
time 
\begin{equation}
t_d=2t_{sd}\frac{\Delta\Omega_c}{\Omega_c},
\end{equation}
as in Equation [\ref{td}]. AAPS
obtain a relaxation time, for perturbations about spin-down equilibrium, of
\begin{equation}
t_r=\frac{\omega_{\rm crit}}{\vert\dot{\Omega}_0\vert}\frac{kT}{E_p} 
=\frac{kT}{\vert\dot\Omega_0\vert\rho_s\kappa r \xi_n a}
\qquad\qquad
\mbox{``non-linear creep''}.
\label{traaps}
\end{equation}
As in this paper, post-glitch response consists of exponential decay
if $t_r>>t_d$ and delayed response if $t_d>>t_r$. 

An important difference between the results of this paper and those of
AAPS is the dependence of the $t_r$ on $E_p$.  In the vortex creep
theory of AAPS, $\omega_{\rm crit}\propto E_p$, giving a relaxation
time that is {\em independent of $E_p$}.  Equation [\ref{traaps}] from
the theory of AAPS gives
\begin{equation}
t_r\simeq 8\, (f_c \rho_{n,14})^{-1}
\left(\frac{\vert\dot\nu\vert}{10^{-11}\mbox{ s$^{-2}$}}\right)^{-1}
\left(\frac{kT}{\mbox{10 keV}}\right)
\left(\frac{\xi_n}{\mbox{10 fm}}\right)^{-1}
\left(\frac{a}{\mbox{50 fm}}\right)^{-1}
\left(\frac{R}{\mbox{10 km}}\right)^{-1}
\mbox{ yr}, 
\end{equation}
far longer than observed, especially if nuclear entrainment with
$f_c<0.1$ is taken into account; 
nuclear entrainment could increase this timescale
to over a century. These timescales are incompatible with 
observations. 

\item In the opposite limit of low equilibrium lag, $\omega_0
<<\omega_{\rm crit}$,
corresponding to relatively small $E_p$, Equations [\ref{valpar}] and
[\ref{Aalpar}] give for the theory of AAPS:
\begin{equation}
\langle\vbf_v\rangle\cdot\hat{r}=2v_0\beta\, E_p{\rm e}^{-\beta
E_p}\frac{\omega}{\omega_{\rm crit}} \qquad \omega_0<<\omega_{\rm
crit}. 
\label{valpar_linear}
\end{equation}
In the nomenclature of
\citet{alpar_etal89}, this limit is the regime of ``linear
creep''. In this regime, post-glitch response consists of 
exponential decay over a timescale that {\em depends exponentially on $E_p$}: 
\begin{equation}
t_r=\frac{r\omega_{\rm crit}}{4\Omega_s v_0}\left(\beta E_p\right)^{-1}{\rm
e}^{\beta E_p}
\qquad\qquad
\mbox{``linear creep''}, 
\label{taulinear}
\end{equation}
and the decoupling time $t_d$ is effectively
zero. Equation [\ref{taulinear}] gives much shorter coupling times than
Equation [\ref{traaps}]; consequently, the regime of
``linear creep'' was invoked by
\citet{alpar_etal93,alpar_etal94,alpar_etal96} to explain observed
post-glitch response. 

\end{enumerate}

The ``linear regime'' of \citet{alpar_etal89} depends on competition
between the two terms in Equation [\ref{valpar}] that gives linear
dependence on $\omega$ for $\omega<<\omega_{\rm crit}$. As shown in
Appendix \ref{antiparallel}, though, thermal activation of vortex segments {\em
against} the Magnus force is prevented by the large vortex tension
unless $\sin 2\theta_d$ is orders of magnitude smaller then
expected. Hence, the second term in Equation [\ref{valpar}] should not be
present, and the ``linear regime'' introduced by
\citet{alpar_etal88,alpar_etal89} is not physically realized. The analysis
of this paper gives, instead, 
\begin{equation}
\langle\vbf_v\rangle\cdot\hat{r}=\frac{1}{2}r\omega\sin 2\theta_d\,
{\rm e}^{-\beta A(\omega)}, 
\end{equation}
for any lag.  The above expression does not have a linear expansion
about $\omega=0$.\footnote{For $\omega<<\omega_{\rm crit}$, the
activation energy diverges as $\omega^{-1}$ \citep{le91}.}

The ``linear creep''
regime was invoked to fit glitch data by 
\citet{alpar_etal93,alpar_etal94,alpar_etal96}. 
In fitting to eight glitches in the Vela pulsar, \citet{alpar_etal93}
included three linear response terms, that is simple exponentials with
no time offsets, with relaxation times of $\sim 30$ d and $\sim 3$ d,
and also 10 hours for the glitch of December 24, 1988. The three
contributions to the linear response are assumed to correspond to
regions through which no vortex motion occurred, and the pinning is
``superweak''. Under this assumption, the change in lag at the glitch
for each region is $\omega_0-\omega(0+)=\Delta\Omega_c$. Assuming
corotation of the core, as did \citet{alpar_etal93}, the decoupling time
for each region follows from Equation [\ref{tt}]. The glitch
magnitudes were in the range $1.14\times 10^{-6}\le
\Delta\Omega_c/\Omega_c\le 3.06\times 10^{-6}$, giving a range of
decoupling times $8\mbox{ d}\le t_d\le 21\mbox{ d}$, 
significantly larger than the relaxation times of 3 d and 10
hours. Hence, the glitches were too large to exhibit exponential recovery
according to the treatment of this paper. 

Whether or not the second term in Equation [\ref{valpar}] is retained, the
creep velocity of AAPS violates the
vortex equations of motion, Equations [\ref{vv}] and [\ref{vslippage}];
Equation [\ref{vslippage}] contains a 
factor of $\omega$ that multiplies the exponential. 
AAPS effectively fixed 
\begin{equation}
\frac{v_s}{2}\sin 2\theta_d=\frac{r\omega}{2}\sin
2\theta_d=v_0=10^7\mbox{ cm s$^{-1}$}, 
\label{v0}
\end{equation}
For a critical lag of $\omega_{\rm crit}=4$ \rads\
(Equation \ref{omc_crust}), $v_0$ cannot exceed $2\times 10^6$ \cms\ for
any drag process; Equation [\ref{v0}] corresponds to unphysically large
drag.  For ``non-linear creep'' the value of $v_0$ is 
not very important, since the assumption of spin-down equilibrium
gives a weak, logarithmic dependence on the prefactor in
Equation [\ref{vslippage}]; see Equation [\ref{trfinal}]. More significant is
the choice of AAPS of the activation energy of Equation [\ref{Aalpar}],
which gives a recoupling time $t_r$ that is independent of of $E_p$ in
this limit, so that the recoupling time depends only on the fact that
there is pinning, and not on the strength of the pinning.
Equation [\ref{Aalpar}] used by
\citet{alpar_etal84a} does not follow from consideration of a
realistic pinning potential or the large vortex self energy. These
considerations give a dependence of the activation energy on
$(1-\omega/\omega_{\rm crit})^{5/4}$ for a parabolic pinning force,
giving $t_r\propto E_p^{9/10}$ for the inner crust and $t_r\propto
E_p^{3/5}$ for the outer core. 

As Equation [\ref{fullsolution}] shows, the response of the crust to a
glitch generally depends non-linearly on the the magnitude of glitch,
through the appearance of $t_d$. Exponential decay occurs only in the
limit $t_d<<t_r$, as for a small glitch. For large glitches,
exponential decay occurs only if $t_r$ exceeds $t_d$. For a large
glitch ($\Delta\nu/\nu=3\times10^{-6}$), the decoupling time is
$21\,(t_{sd}/10^4\mbox{ yr})$ d; in regions with pinning, {\em
exponential decay can occur only over longer timescales than the
decoupling time.}

\citet{sa09} studied response of the outer core assuming 
``linear creep'', a regime that is not realized for physically
reasonable parameters 
(see Appendix \ref{antiparallel}). With Equation [\ref{valpar_linear}],
\citet{sa09} 
obtained a sub-day relaxation time assuming $E_p\simeq 0.1$ MeV,
much smaller than the estimate of 100 MeV discussed in Section
\ref{pinning}. From the analysis of Appendix \ref{transition}, pinning
does not exist for such a low value of $E_p$. For larger values of
$E_p$ in the pinning regime, $t_r$ can be short (\eg, $E_p=1$ MeV
gives $t_r=3$ hours for Vela), but $t_d$ will still be approximately
one week for $\Delta\nu/\nu=10^{-6}$.

\section{Conclusions}

\label{predictions}

Two chief conclusions of this paper are:

\begin{enumerate}

\item 
The ``linear
creep'' regime introduced by
\citet{alpar_etal88,alpar_etal89} and invoked in fits to post glitch
response \citep{alpar_etal93,alpar_etal94,alpar_etal96} is not
realized for physically reasonable parameters, a conclusion that
strongly constrains possible post-glitch response through thermal
activation. The validity of fits of vortex creep theory to data, and
the conclusions drawn about pinning parameters, should be reassessed.

\item 
A regime of ``superweak pinning'', crucial to the theory of
AAPS and its extensions, is probably precluded by thermal
fluctuations. 

\end{enumerate}

The vortex slippage theory given here has clear, falsifiable predictions:

\begin{enumerate}

\item Post-glitch recovery associated with regions in which there is
vortex pinning, whether in the inner crust or outer core, requires a
time that is {\em at least} as long as the decoupling time:
\begin{equation}
\mbox{Recovery time } > 
t_d\simeq 7\,\left(\frac{t_{sd}}{10^4\mbox{ yr}}\right)
\left(\frac{\Delta\nu/\nu}{10^{-6}}\right)
\mbox{ days}.
\end{equation}
This timescale is independent of the details of vortex pinning, and so
is a robust conclusion within the framework of thermally-activated,
post-glitch response. 

\item Post-glitch response depends non-linearly on the size of the
glitch. For a glitch that is sufficiently large that 
$t_d>>t_r$ is satisfied, the response will consist of 
{\em delayed response} over a time $t_d$, wherein the star {\em spins
down at a constant, higher rate, before relaxing over a time $t_r$}. 
Detection of delayed response
over a timescale proportional to the glitch magnitude would represent
strong evidence for thermally-activated vortex slippage in neutron
stars. 

\item Exponential decay can occur only if the glitch is sufficiently
small that $t_d<<t_r$ is satisfied. Hence,
exponential decay can occur only over timescales that exceed $t_d$,
which is determined by the size of the glitch. Quicker decay is
prevented by the non-linear nature of coupling through thermal
activation. This conclusion is an important difference between vortex
slippage theory and vortex creep theory. 

\item Response over a relatively quick timescale $<t_d$ 
as has been observed in some pulsars, most notably Vela, 
cannot be due to thermally-activated vortex motion but must
represent a different process, such as drag on vortices in regions
where there is no pinning.

\end{enumerate}

Comparison of these predictions with the wealth of glitch data now
available will be given in a forthcoming publication, and constraints
upon the theory will be obtained. 

\begin{acknowledgments}
I am grateful to M. A. Alpar, D. Antonopoulou, N. Chamel, and
C. Espinoza for very useful discussions, and to R. I. Epstein for
valuable suggestions on the manuscript. I also thank the anonymous
referee for very useful criticism. This work was supported by NSF
Award AST-1211391 and NASA Award NNX12AF88G.
\end{acknowledgments}

\appendix

\section{Transition from Vortex Pinning to Vortex Drag}

\label{transition}

For a sufficiently weak pinning interaction, the thermal energy per
unit length of a vortex will exceed the pinning interaction per unit
length. In this limit the concept of pinning is no longer appropriate;
vortex motion occurs against a drag force, rather than against pinning
barriers. The transition occurs when the pinning energy per unit
length $E_p/l_p$ becomes less than the thermal energy per unit
length $E_T/L$ of a free vortex. 

Consider a free vortex in thermal equilibrium with its environment at
temperature $T$. The equation of motion of the vortex of amplitude
$\ubf(z,t)$ is \citep{eb92}
\begin{equation}
T_v\frac{\partial^2\ubf(z,t)}{\partial z^2}
+\rho_s\,\kappabf\times\frac{\partial\ubf(z,t)}{\partial t}=0.
\end{equation}
The solutions at the quantum level are ``kelvons'', 
right and left circularly polarized waves with energies
\begin{equation}
\epsilon_k=\frac{(\hbar k)^2}{2\mu}, 
\end{equation}
where $k$ is the wavenumber and $\mu\equiv\rho_s\kappa\hbar/2T_v$ is
the kelvon effective mass. Typically $\mu\simeq m_n/3\pi$ where $m_n$ is
the neutron mass; see \citet{eb92}. 

The kelvons behave as a one-dimensional gas of bosons, with two
polarization states and zero chemical potential. The thermal energy
per unit length is 
\begin{equation}
\frac{E_T}{L}=2\int_0^\infty \frac{dk}{2\pi}\,\frac{\epsilon_k}{{\rm
e}^{\beta\epsilon_k}-1} = 
0.52\,\frac{\sqrt{\mu}}{\hbar}(kT)^{3/2}.
\end{equation}
Evaluating $l_p$ with Equation [\ref{lp}] for the inner crust 
gives $E_p/l_p<E_T/L$ for $E_p$ in the range
\begin{equation}
E_p<0.04\mbox{ MeV}\,
\left(\frac{kT}{\mbox{10 keV}}\right)^{3/2}
\left(\frac{a}{\mbox{50 fm}}\right)
\left(\frac{f_c}{0.1}\right)^{1/3}
\left(\frac{\rho_{n,14}}{0.5}\right)^{1/3}
\qquad \mbox{(inner crust)}.
\label{Ep_trans_crust}
\end{equation}
Using $l_p$ with Equation [\ref{lphi}] for the outer core gives
$E_p/l_p<E_T/L$ for 
\begin{equation}
E_p<0.3\mbox{ MeV}\,
\left(\frac{kT}{\mbox{10 keV}}\right)^{3/2}B_{12}^{-1/2}
\qquad \mbox{(outer core)}.
\end{equation}
For interaction energies below these upper limits, vortices do not
pin. 

\appendix

\setcounter{section}{1}

\section{Vortex Motion Anti-Parallel to the Magnus Force}

\label{antiparallel}

Equation [\ref{vslippage}] accounts for thermally-activated motion from
pinning location $P_1$ to $P_2$; see Figure \ref{trajectories}. Motion to
point $P^\prime$, anti-parallel to the Magnus force, is strongly
forbidden by vortex tension.

Suppose a segment of vortex of length $L_p=N_pl_p$ unpins from pinning point
$P_1$, moves under dissipation at angle $\theta_d$ with respect to
$\hat{\phi}$, and repins at a single pinning point $P_2$ a distance
$d\sin\theta_d$ further from the rotation axis. To reach $P_2$, the
vortex must unzip to a final length $L_f=4T_vdr_0/E_p$
(Equation \ref{Lf}). 

A competing process is for the vortex to go the point $P^\prime$,
against the Magnus force. After unpinning and unzipping, the vortex
can pin at $P^\prime$ (the condition for mechanical equilibrium
is the same as for $P_2$), but must increase its energy through
thermal fluctuations by
\begin{equation}
\Delta E=2f_{\rm mag}L_fd\sin\theta_d=8f_c\rho_n\kappa T_v E_p^{-1}
\, d^2rr_0\,\omega\sin\theta_d.
\label{dE}
\end{equation}
The energy penalty is lowered for low lag and low dissipation, but
{\em increases} with decreasing $E_p$, because $L_f$ becomes larger. 
Transitions to $P^\prime$, by any path, are relevant only if 
\begin{equation}
\beta\Delta E\lap 1. 
\end{equation}
To obtain $\Delta E$, $\omega$ is obtained 
from the solution 
\begin{equation}
\langle\vbf_v\rangle\cdot\hat{r}=r\omega\sin 2\theta_d\,{\rm
e}^{-\beta A}
- r\omega\sin 2\theta_d\,{\rm
e}^{-\beta (A+\Delta E)}=
\frac{r}{4t_{sd}}. 
\label{sdeq}
\end{equation}
The first term accountsfor motion to $P_2$, while the second term
accounts for motion to $P^\prime$. 
Numerical solution to the above equation using the activation energy
$A$ from eq [\ref{A}], for typical parameters,  gives
$\omega_0\simeq\omega_c$. 
The condition $\beta\Delta E<1$ gives, for the inner crust, with $d=a$
and $t_{sd}=10^4$ yr, 
\begin{equation}
\sin 2\theta_d< 1.5\times 10^{-4}\, 
E_p(\mbox{MeV})^{-1/2}\left
(\frac{f_c\rho_{n,14}}{0.5}\right)^{-1/2}
\left(\frac{a}{\mbox{50 fm}}\right)^{-1/2}
\left(\frac{kT}{\mbox{10 keV}}\right)
\qquad
\mbox{(inner crust)}
\end{equation}
Transitions to $P^\prime$ are important only if $\sin 2\theta_d$ is
smaller than this limit, and are otherwise suppressed by a factor
${\rm e}^{-\beta\Delta E}$ relative to transitions to $P_2$. In the
inner crust, $\sin 2\theta_d\gap 10^{-2}$ is expected.  Consider
negligible nuclear entrainment, $E_p=0.1$ MeV (the lower limit for
pinning), $kT=10$ keV, $\rho_{n,14}=0.5$, and $\sin 2\theta_d\ge
10^{-2}$. Transitions to $P^\prime$ are suppressed relative to
transitions to $P_2$ by a typical factor $\le {\rm e}^{-15}$, with
greater suppression for higher $E_p$. With nuclear entrainment
($f_c=0.1$), $E_p=0.04$ MeV (the lower limit for pinning), 
$\rho_{n,14}=0.5$, and $\sin 2\theta_d\ge 10^{-2}$, the suppression
factor is $\le {\rm e}^{-1.5}\simeq 0.22$, with greater suppression at
higher $E_p$.

In the outer core, $\sin 2\theta_d\gap 10^{-3}$ is expected. Here
$d=l_\Phi$. Taking $\rho_{n,14}=4$, transitions to $P^\prime$ are
suppressed relative to transitions to $P_2$ by a
typical factor $\le {\rm e}^{-5\times 10^4}$, independent of $E_p$. 
The condition $\beta\Delta E<1$ is met if 
\begin{equation}
\sin 2\theta_d>2\times 10^{-8}\, 
\left(\frac{\rho_{n,14}}{4}\right)^{-1}
\left(\frac{kT}{\mbox{10 keV}}\right)
B_{12}^{1/2}
\qquad
\mbox{(outer core)}. 
\end{equation}

Hence, in general, transitions against $\fbf_{mag}$ are 
strongly suppressed by the effects of vortex tension in both the inner
crust and the outer core, and are effectively
excluded for reasonable values of $\sin 2\theta_d$. 

\appendix

\setcounter{section}{2}

\section{Random Pinning}

\label{random_pinning}

At baryon densities above $\rho_b\simeq 6.3\times 10^{13}$ \gcc\ in
the inner crust, the
coherence length $\xi_n$ begins to exceed the Wigner-Seitz spacing
\citep{schwenk_etal03,cao_etal06}. 
A vortex now encompasses many pinning nuclei, and pinning will be
greatly weakened. \citet{alpar_etal84a} refer to this regime as
``superweak pinning''. A vortex now pins to spatial variations in the
number density of pinning potentials; I refer to this regime as
``random pinning''.  Here I show that random pinning is sufficiently
weak that it might be precluded by thermal fluctuations.

Consider a vortex of radius $\xi_n>>a$ containing {\rm randomly
distributed} point-like potentials of spacing $a$ and number density
$n_p=a^{-3}$. The energy associated with the overlap of the vortex
with $N$ randomly placed potentials of strength $E_p$ is $E\simeq
N^{1/2}E_p$. The average interaction energy per potential is $\langle
E_p\rangle= N^{-1/2}E_p$.

A vortex segment can be regarded as interacting with $N$ nuclei in a
flat disc of radius $\xi_n$, thickness $a$, and volume $V=\pi\xi_n^2
a$, containing $N=n_pV$ interaction nuclei, giving
$N=n_p\pi\xi_n^2a$. A rough estimate of the effective pinning energy
per nucleus in this regime is thus
\begin{equation}
\langle E_p\rangle\sim \frac{a}{\sqrt{\pi}\xi_n}E_p.
\end{equation}
The effective pinning energy increases as 
superfluidity weakens, and $\xi_n$ becomes large. 

\citet{chamel05,chamel12} finds that nuclear entrainment is significant
from just above neutron drip to densities somewhat below
$\rho_{b,14}=1$, where the effects become relatively small. 
For random pinning to occur, $\langle E_p\rangle$ must exceed the lower
limit given by [\ref{Ep_trans_crust}] at which thermal excitations
preclude pinning, giving
\begin{equation}
E_p>0.4\, \left(\frac{kT}{10\mbox{ keV}}\right)^{3/2}
\left(\frac{\xi_n}{100\mbox{ fm}}\right)\mbox{ MeV}, 
\end{equation}
where a value of $\xi_n$ appropriate to $\rho_{b,14}=1$ was used, and
$f_c$ was taken to be unity. 
Pinning calculations show a sharp drop in $E_p$ to
nearly zero above $\rho_{b,14}\simeq 0.6$ \citep{dp06,abbv07}.

These arguments should not be significantly altered in the pasta
regime at higher densities. Here $\xi_n$ is predicted to greatly
exceed the cell spacing, and the detailed shapes of the nuclear
clusters are unlikely to play an important role in 
the basic energetics. 

These considerations suggest that random or ``superweak'' pinning
probably does not occur at all for $\rho_b\gap 6\times 10^{13}$ \gcc,
from the denser regions of the inner crust into the pasta region.


\end{document}